\documentclass
[showpacs,amsfonts,amssymb,oneside,notitlepage,10pt,a4paper,twocolumn,preprint,balancelastpage,pra]{revtex4}%
\usepackage{amsfonts}
\usepackage{amsmath}
\usepackage{amssymb}
\usepackage{graphicx}
\usepackage[nonscript]{accents}%
\providecommand{\U}[1]{\protect\rule{.1in}{.1in}}

\begin{document}
\title{Non-critical squeezing in 2-transverse-mode optical parametric oscillators}
\author{Carlos Navarrete-Benlloch$^{1}$, Alejandro Romanelli$^{2}$, Eugenio
Rold\'{a}n$^{1}$ and Germ\'{a}n J. de Valc\'{a}rcel$^{1}$.}
\affiliation{$^{1}$Departament d'\`{O}ptica, Universitat de Val\`{e}ncia, Dr. Moliner 50,
46100--Burjassot, Spain.
\\
$^{2}$Instituto de F\'{\i}sica, Facultad de Ingenier\'{\i}a Universidad de la
Rep\'{u}blica, C.C. 30, C.P. 11000, Montevideo, Uruguay.}

\begin{abstract}
In this article we explore the quantum properties of a degenerate optical
parametric oscillator when it is tuned to the first family of transverse modes
at the down converted frequency. Recently we found [Phys. Rev. Lett.
\textbf{100}, 203601 (2008)] that above threshold a TEM$_{10}$ mode following
a random rotation in the transverse plane emerges in this system (we denote it
as \textit{bright} mode), breaking thus its rotational invariance. Then, owed
to the undetermination of the mode orientation, we showed that the phase
quadrature of the transverse mode orthogonal to this one (denoted as
\textit{dark} mode) is perfectly squeezed at any pump level and without an
increasing of the fluctuations on its amplitude quadrature (which seems to
contradict the uncertainty principle). In this article we go further in the
study of this system and analyze some important features not considered
previously. First we show that the apparent violation of the uncertainty
principle is just that, apparent, as the conjugate pair of the squeezed
quadrature is not another quadrature but the orientation of the bright mode
(which is completely undetermined in the long term). We also study an homodyne
scheme in which the local oscillator is not perfectly matched to the dark
mode, as this could be impossible in real experiments due to the random
rotation of the mode, showing that even in this case large levels of noise
reduction can be obtained (also including the experimentally unavoidable phase
fluctuations). Finally we show that neither the adiabatic elimination of the
pump variables nor the linearization of the quantum equations are responsible
for the remarkable properties of the dark mode (things which we prove
analytically and through numerical simulations respectively), which were
simplifying assumptions used in [Phys. Rev. Lett. \textbf{100}, 203601
(2008)]. These studies show that the production of non-critically squeezed
light through the spontaneous rotational symmetry breaking is a robust phenomenon.

\end{abstract}

\maketitle

\section{Introduction}

One of the most amazing predictions offered by the quantum theory of light is
what has been called \textit{vacuum fluctuations}: Even in absence of photons
(vacuum) the value of the fluctuations of some observables are different from
zero. These fluctuations cannot be removed by improving the experimental
instrumental, and hence, they are a source of non-technical noise
(\textit{quantum noise}) which seems to establish a limit for the precision of
experiments involving light.

During the late seventies and mid-eighties of the past century ways for
overcoming this fundamental limit were predicted and experimentally
demonstrated \cite{KeyPapers}. In the case of the quadratures of light
(equivalent to the position and momentum of a harmonic oscillator), the trick
was to eliminate (\textit{squeeze}) quantum noise from one quadrature at the
expense of increasing the noise of its canonically conjugated one, in order to
preserve their Heisenberg uncertainty relation. States with this property are
called \textit{squeezed states}, and they can be generated by means of
nonlinear optical processes. In particular, the largest levels of squeezing
are obtained by using nonlinear resonators operating near threshold, this
level being degraded quickly as one moves away from that critical point.

Up to date, the best squeezing ever achieved is a 90\% of noise reduction with
respect to vacuum (the so-called \textit{standard quantum limit})
\cite{Vahlbruch08,Takeno07}, and \textit{degenerate optical parametric
oscillators} (DOPOs) have been used to this aim. A DOPO consists on a
nonlinear $\chi^{\left(  2\right)  }$ crystal placed inside an optical cavity;
when pumped above some threshold level with a laser beam at frequency
$2\omega_{0}$, a field oscillating at half that frequency (\textit{signal}
frequency) is generated. A linearized analysis of the DOPO's quantum
fluctuations reveals that the phase quadrature of the signal field can be
perfectly squeezed when pump's power is tuned to the DOPO's threshold value
\cite{Collett84}. Of course, ideal perfect squeezing cannot be real in this
case, as it would entail an infinite number of photons in the generated mode,
and it can be shown that nonlinear corrections make this squeezing level
become finite \cite{NonlinearDOPO}.

Apart from fundamental reasons, improving the quality of squeezed light is an
important task because of its applications. For example, in the fields of
\textit{quantum information with continuous variables }\cite{QICV} (as mixing
squeezed beams with beam splitters offers the possibility to generate
multipartite entangled beams \cite{vanLoock00Aoki03}) and \textit{high}
\textit{precision measurements }(such as beam displacement and
pointing\textbf{ }measurements \cite{Treps}\ or gravitational wave detection
\cite{Vahlbruch05Goda08}), applications of squeezed light have been
theoretically and experimentally proved.

Recently, our group has developed a strategy that would lead to the generation
of light with high level of squeezing at any pump level above threshold in
DOPOs (\textit{non-critically squeezed light}). The idea is to allow for the
existence of several transverse modes at the signal frequency in a DOPO
possessing some spatial symmetry in the transverse plane (e.g., rotational or
translational); under some circumstances, a pattern breaking the corresponding
spatial symmetry can be generated, and quantum noise can randomly move it
along the invariant direction (i.e., rotate or translate it in the transverse
plane). This can be seen as an indefiniteness in the transverse position of
the generated pattern, which invoking now the uncertainty principle, could be
accompanied by the perfect determination of its associated momentum (angular
or linear). Hence, we could expect noise reduction in the empty pattern
coinciding with the momentum of the generated pattern. Moreover, as this is
related to the spontaneous symmetry breaking, the perfect squeezing should
occur at any pump level above threshold.

In Ref. \cite{EPL+PRA}, the translational symmetry breaking was considered in
broad area, planar DOPOs, where cavity solitons (CSs), as well as extended
patterns, have been predicted to exist \cite{CS}. It was shown that all the
reasoning above is true: The position of the CS diffuses in the transverse
plane, and the phase quadrature of the pattern coinciding with its linear
momentum (namely its $\pi/2$ phase shifted transverse gradient) is perfectly
squeezed (within the linearized theory) at any pump level above threshold. The
problem of this model is that it is not too close to current experimental
setups and even CSs have not yet been observed in DOPOs.

This was one of the reasons why the theory of a simpler system was developed
in Ref. \cite{PRL}: A rotationally symmetric DOPO tuned to the first family of
transverse modes at the signal frequency \cite{BTExperiments,EXP}. The first
transverse-mode family supports Laguerre-Gauss (LG) modes $L_{\pm1}\left(
\mathbf{r}\right)  $ with $\pm1$ orbital angular momentum (OAM), where
$\mathbf{r}=\left(  x,y\right)  $ are the transverse coordinates. When pumped
with a Gaussian mode with zero OAM, two signal photons with opposite OAM are
generated in the $\chi^{\left(  2\right)  }$ crystal. Alternatively, the
simultaneous generation of one $L_{+1}\left(  \mathbf{r}\right)  $ photon and
another $L_{-1}\left(  \mathbf{r}\right)  $ one corresponds to the generation
of two Hermite-Gauss (HG) TEM$_{10}$ photons, the orientation of this
TEM$_{10}$ mode ($\theta$ in Fig. 1) given by half the phase difference
between the subjacent $L_{\pm1}\left(  \mathbf{r}\right)  $ modes. Again in
this case, we proved our reasoning given above: Quantum noise is able to
rotate randomly this bright mode, and the phase quadrature of its angular
momentum (its $-\pi/2$ phase shifted angular derivative), which corresponds in
this case to another HG mode orthogonal to the generated one (we will refer to
it as the dark mode), is perfectly squeezed within the linearized theory and
at any pump level above threshold \cite{OPOnote,OPO}. In addition we showed
that this result is quite robust against deviations from the perfect
rotational invariance of the DOPO.%

\begin{figure}[ht]

\includegraphics[
width=3.4in
]%
{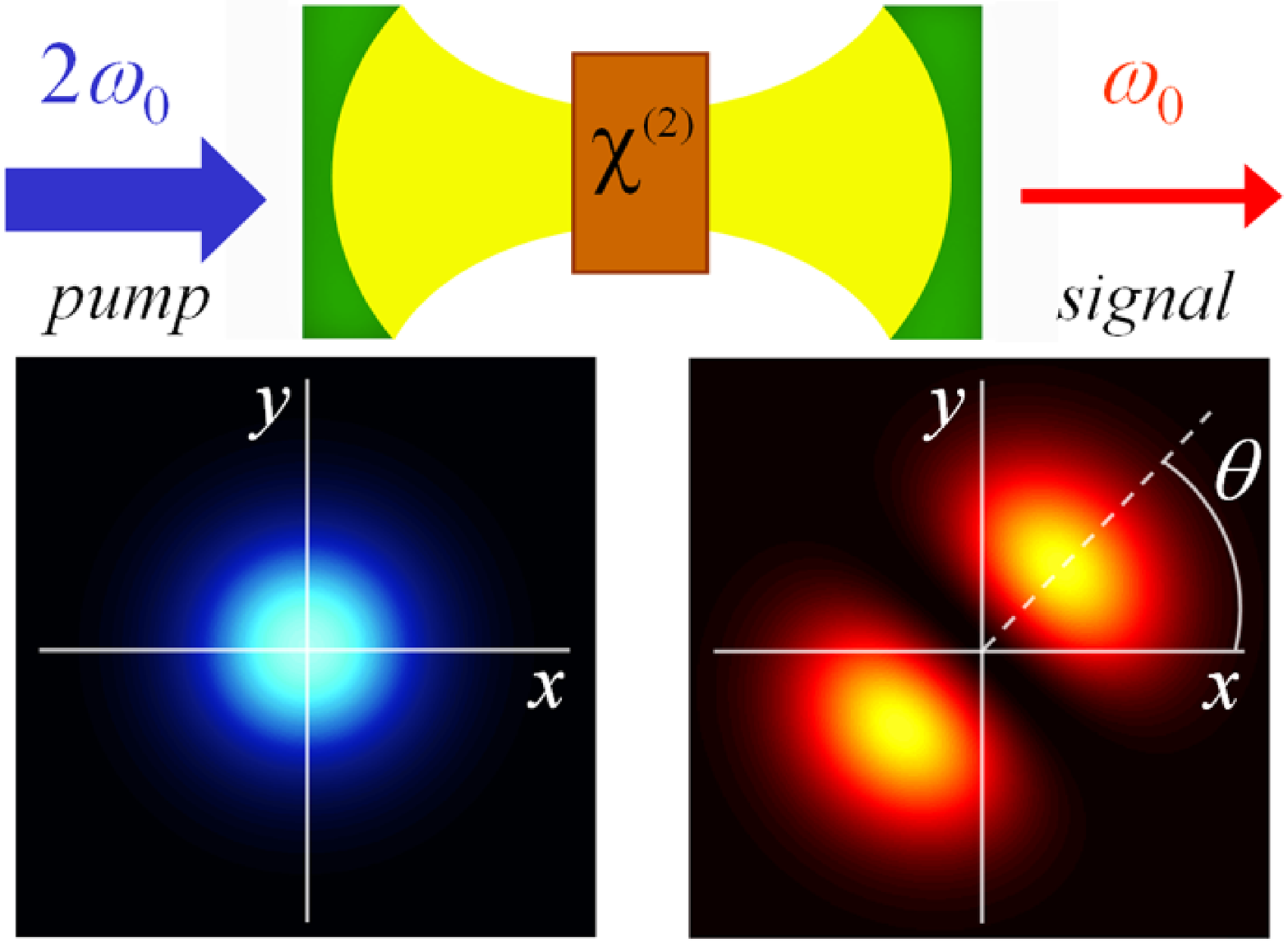}%
\\
{\small Figure 1.- Scheme of the 2-transverse-mode DOPO.}%

\end{figure}

Moreover, a surprising result was found: although the phase quadrature of the
dark mode is perfectly squeezed, its corresponding amplitude quadrature has
not increased uncertainty, i.e., the quadratures of this rotating mode do not
form a canonical pair. We intuitively explained this by noting that the excess
of noise is transferred not to another quadrature, but to the orientation
$\theta$ of the pattern, which actually is the canonical pair of the squeezed
quadrature; in the present article we prove this statement by using the
arguments developed in Ref. \cite{arXiv}.

Indeed, the study of this rotationally symmetric DOPO was carried out under
some assumptions whose repercussion will be analyzed in the current article.
First, we supposed that cavity losses at the signal frequency were small as
compared with that at the pump frequency, what allowed us to adiabatically
eliminate the pump variables. Next, the usual linearization of the nonlinear
Langevin equations around their classical stationary solution was done.
Finally, in order to prove the non-critical squeezing we proposed an homodyne
detection scheme in which the local oscillator was a TEM$_{10}$ mode matched
to the dark mode at any time, i.e., it was following the random rotation of
this mode, an ideal situation which is not possible in real experiments.

Now, let us summarize what we are going to show in this article, which is
divided into four main parts. In the first one (Section II) we describe the
DOPO tuned to the first transverse family (the \textit{2-transverse-mode
DOPO}) and find its associated quantum Langevin equations. In Section III, we
study the classical emission of the system, and show how the rotational
symmetry is broken by the generated transverse pattern. Section IV is the main
section of the article: The quantum properties of the DOPO are analyzed by
linearizing the Langevin equations. Here we will prove the random rotation of
the classical pattern, and find the squeezing properties of the bright and
dark modes. In addition, we will show that the quadratures of the dark mode do
not form a canonical set, but they are canonically related to the pattern
orientation (where the excess of noise goes). We close this section by
analyzing the repercussion of keeping fixed the orientation of the local
oscillator, showing that even in this case large levels of squeezing can be
obtained. In the last part (Section V) we show that the approximations
considered in the previous sections (the adiabatic elimination of the pump and
the linearization procedure) are not responsible for the remarkable squeezing
properties of the dark mode, as they follow directly from the spontaneous
rotational symmetry breaking. At the end (Section VI) we will give some
conclusions. In order to make clearer the physics behind the phenomenon, we
have left most of the technical details to appendices.

\section{Description of the 2-transverse mode DOPO}

\subsection{The field inside the cavity}

One of the simplest configurations of the DOPO's cavity allowing for the
generation of a rotationally asymmetric pattern is the following \cite{PRL}:
At some phase-matched frequency $2\omega_{0}$ it is tuned to a fundamental
Gaussian transverse mode $G\left(  \mathbf{r}\right)  $, while at half that
frequency, $\omega_{0}$, the first transverse family is resonant. In the first
transverse family two Laguerre-Gauss modes exist, $L_{\pm1}\left(
\mathbf{r}\right)  $. The exact expression of the modes in the waist plane of
the resonator (where the $\chi^{\left(  2\right)  }$ crystal is placed) is
given by \cite{Hodgson05}%
\begin{align}
G\left(  \mathbf{r}\right)   &  =\sqrt{\frac{2}{\pi}}\frac{1}{w_{\mathrm{p}}%
}e^{-\frac{r^{2}}{w_{\mathrm{p}}^{2}}}\\
L_{\pm1}\left(  \mathbf{r}\right)   &  =\frac{2}{\sqrt{\pi}}\frac
{r}{w_{\mathrm{s}}^{2}}e^{-\frac{r^{2}}{w_{\mathrm{s}}^{2}}}e^{\pm i\phi
},\nonumber
\end{align}
where $\mathbf{r}=r\left(  \cos\phi,\sin\phi\right)  $ is the coordinate
vector in the transverse plane, and $w_{i}$ is the beam radius at the waist
plane, which depends on the cavity geometry and the resonance frequency of the
mode (see Appendix A).

These transverse modes have $0$ and $\pm1$ OAM respectively, as they are
eigenmodes of the OAM operator $-i\partial_{\phi}$ with those eigenvalues.

The field inside the cavity can be written then as%
\begin{equation}
\hat{E}=i\mathcal{F}_{\mathrm{p}}\hat{A}_{\mathrm{p}}\left(  \mathbf{r}%
,t\right)  e^{-2i\omega_{0}t}+i\mathcal{F}_{\mathrm{s}}\hat{A}_{\mathrm{s}%
}\left(  \mathbf{r},t\right)  e^{-i\omega_{0}t}+\mathrm{H.c.},
\end{equation}
with%
\begin{equation}
\mathcal{F}_{\mathrm{p}}=\sqrt{2}\mathcal{F}_{\mathrm{s}}=\sqrt{\frac
{2\hbar\omega_{0}}{n\varepsilon_{0}L}},
\end{equation}
being $n$ the refractive index of the crystal and $L$ the effective cavity
length (H.c. stands for \textquotedblleft Hermitian
conjugate\textquotedblright). The slowly varying envelopes are given by
\begin{subequations}
\label{SVE}%
\begin{align}
\hat{A}_{\mathrm{p}}\left(  \mathbf{r},t\right)   &  =\hat{a}_{0}\left(
t\right)  G\left(  \mathbf{r}\right) \label{SVEp}\\
\hat{A}_{\mathrm{s}}\left(  \mathbf{r},t\right)   &  =\hat{a}_{+1}\left(
t\right)  L_{+1}\left(  \mathbf{r}\right)  +\hat{a}_{-1}\left(  t\right)
L_{-1}\left(  \mathbf{r}\right)  , \label{SVEs}%
\end{align}
where the boson operators satisfy the usual equal time commutation relations%
\end{subequations}
\begin{equation}
\left[  \hat{a}_{m}\left(  t\right)  ,\hat{a}_{n}^{\dagger}\left(  t\right)
\right]  =\delta_{mn};\text{ }m,n=0,\pm1.
\end{equation}

Instead of using the LG basis, one can work in the most usual TEM$_{mn}$ (HG)
basis. Denoting by $H_{10}^{\psi}\left(  \mathbf{r}\right)  $ and
$H_{01}^{\psi}\left(  \mathbf{r}\right)  $ a TEM$_{10}$ mode rotated an angle
$\psi$ with respect to the $x$ axis and its orthogonal, these are given by
\begin{subequations}
\label{HG}%
\begin{align}
H_{10}^{\psi}\left(  \mathbf{r}\right)   &  =\frac{1}{\sqrt{2}}\left[
e^{-i\psi}L_{+1}\left(  \mathbf{r}\right)  +e^{i\psi}L_{-1}\left(
\mathbf{r}\right)  \right]  \label{HG10}\\
&  =\sqrt{2}\left\vert L_{\pm1}\left(  \mathbf{r}\right)  \right\vert
\cos\left(  \phi-\psi\right)  \nonumber\\
H_{01}^{\psi}\left(  \mathbf{r}\right)   &  =\frac{1}{\sqrt{2}i}\left[
e^{-i\psi}L_{+1}\left(  \mathbf{r}\right)  -e^{i\psi}L_{-1}\left(
\mathbf{r}\right)  \right]  \label{HG01}\\
&  =\sqrt{2}\left\vert L_{\pm1}\left(  \mathbf{r}\right)  \right\vert
\sin\left(  \phi-\psi\right)  .\nonumber
\end{align}

One can thus define boson operators associated to these modes, whose relation
with the LG ones is
\end{subequations}
\begin{subequations}
\label{LGtoHG}%
\begin{align}
\hat{a}_{10,\psi}  &  =\frac{1}{\sqrt{2}}\left(  e^{i\psi}\hat{a}%
_{+1}+e^{-i\psi}\hat{a}_{-1}\right) \label{LGtoHGc}\\
\hat{a}_{01,\psi}  &  =\frac{i}{\sqrt{2}}\left(  e^{i\psi}\hat{a}%
_{+1}-e^{-i\psi}\hat{a}_{-1}\right)  . \label{LGtoHGs}%
\end{align}
These relations will be important in order to understand the properties of the system.

Finally, note that for any mode $m$, quadrature $\hat{X}_{m}^{\varphi}$ can be
defined as%
\end{subequations}
\begin{equation}
\hat{X}_{m}^{\varphi}=e^{-i\varphi}\hat{a}_{m}+e^{i\varphi}\hat{a}%
_{m}^{\dagger}, \label{Quadrature}%
\end{equation}
and quadratures with $\varphi=0$ and $\varphi=\pi/2$ are usually called
amplitude and phase quadratures, and will be denoted by $\hat{X}$ and $\hat
{Y}$, respectively.

\subsection{Model equations of the system}

Once the structure of the field inside the cavity has been described, we can
pass to describe its evolution, and to this aim a Hamiltonian must be built.
It has to take into account two processes: the pumping of the coherent,
Gaussian, resonant laser source at frequency $2\omega_{0}$, and the parametric
down conversion of the pump photons into signal photons occurring in the
$\chi^{\left(  2\right)  }$ crystal. In the interaction picture, this
Hamiltonian is given by%
\begin{equation}
\hat{H}=i\hbar\left(  \mathcal{E}_{\mathrm{p}}\hat{a}_{0}^{\dagger}+\chi
\hat{a}_{0}\hat{a}_{+1}^{\dagger}\hat{a}_{-1}^{\dagger}\right)  +\mathrm{H.c.}%
, \label{Hamiltonian}%
\end{equation}
where $\mathcal{E}_{\mathrm{p}}$ and $\chi$ are proportional to the injected
pump amplitude and to the second order nonlinear susceptibility of the
nonlinear crystal respectively. Explicit expressions of these parameters in
terms of physical quantities are given in Appendix A. We take $\mathcal{E}%
_{\mathrm{p}}$ as real, i.e., we take the phase of the pump laser as the
reference phase for all the other fields.

The first part of the Hamiltonian is the usual pump Hamiltonian
\cite{GardinerZoller}, while the second part is easily justified by energy and
OAM conservation: From one $2\omega_{0}$ photon with zero OAM, two $\omega
_{0}$ photons are created with opposite OAM, or viceversa. Note that this
Hamiltonian is equivalent to that for the nondegenerate optical parametric
oscillator considered for example in Refs. \cite{McNeil83, OPO}.

There exists one more process not taken into account in this Hamiltonian: The
losses through the cavity mirrors. This is an irreversible process which
cannot be described with a Hamiltonian formalism that takes into account only
the cavity dynamics. However, assuming that the outer of the cavity consists
of a continuum of modes in a vacuum (or coherent) state which weakly interact
with the internal modes through the partially reflecting mirror, it can be
incorporated into the master equation satisfied by the density operator of the
system as shown for example in Ref. \cite{Carmichael99}. Moreover, the master
equation can be converted into a set of stochastic (Langevin) differential
equations by using a positive \textit{P} representation for the density
operator \cite{Drummond80}. In our case, as is shown in \cite{McNeil83}, the
equivalent set of Langevin equations is%
\begin{align}
\dot{\alpha}_{0}  &  =\mathcal{E}_{\mathrm{p}}\mathcal{-}\gamma_{\mathrm{p}%
}\alpha_{0}-\chi\alpha_{+1}\alpha_{-1}\label{Langevin}\\
\dot{\alpha}_{0}^{+}  &  =\mathcal{E}_{\mathrm{p}}\mathcal{-}\gamma
_{\mathrm{p}}\alpha_{0}^{+}-\chi\alpha_{+1}^{+}\alpha_{-1}^{+}\nonumber\\
\dot{\alpha}_{+1}  &  =-\gamma_{\mathrm{s}}\alpha_{+1}+\chi\alpha_{0}%
\alpha_{-1}^{+}+\sqrt{\chi\alpha_{0}}\xi\left(  t\right) \nonumber\\
\dot{\alpha}_{+1}^{+}  &  =-\gamma_{\mathrm{s}}\alpha_{+1}^{+}+\chi\alpha
_{0}^{+}\alpha_{-1}+\sqrt{\chi\alpha_{0}^{+}}\xi^{+}\left(  t\right)
\nonumber\\
\dot{\alpha}_{-1}  &  =-\gamma_{\mathrm{s}}\alpha_{-1}+\chi\alpha_{0}%
\alpha_{+1}^{+}+\sqrt{\chi\alpha_{0}}\xi^{\ast}\left(  t\right) \nonumber\\
\dot{\alpha}_{-1}^{+}  &  =-\gamma_{\mathrm{s}}\alpha_{-1}^{+}+\chi\alpha
_{0}^{+}\alpha_{+1}+\sqrt{\chi\alpha_{0}^{+}}\left[  \xi^{+}\left(  t\right)
\right]  ^{\ast},\nonumber
\end{align}
where $\alpha_{m}$ and $\alpha_{m}^{+}$ are independent complex amplitudes,
$\gamma_{\mathrm{p}}$ and $\gamma_{\mathrm{s}}$ are the decay rates of the
cavity at the pump and signal frequencies respectively (see Appendix A), and
the independent complex noises $\xi$ and $\xi^{+}$ have zero mean and nonzero
correlations%
\begin{equation}
\left\langle \xi\left(  t_{1}\right)  \xi^{\ast}\left(  t_{2}\right)
\right\rangle =\left\langle \xi^{+}\left(  t_{1}\right)  \left[  \xi
^{+}\left(  t_{2}\right)  \right]  ^{\ast}\right\rangle =\delta\left(
t_{1}-t_{2}\right)  . \label{NoiseCorr}%
\end{equation}

The equivalence between the master equation of the system and these equations
must be understood in the following way:%
\begin{equation}
\left\langle :f\left(  \hat{a}_{m},\hat{a}_{m}^{\dagger}\right)
:\right\rangle =\left\langle f\left(  \alpha_{m},\alpha_{m}^{+}\right)
\right\rangle _{\mathrm{stochastic}}, \label{N-St}%
\end{equation}
i.e., quantum expected values of normally ordered functions equal stochastic
averages of the same functions changing boson operators $\left(  \hat{a}%
_{m},\hat{a}_{m}^{\dagger}\right)  $ by independent complex stochastic
variables $\left(  \alpha_{m},\alpha_{m}^{+}\right)  $. Note that these
stochastic equations are equal in either the Ito or Stratonovich forms (see
Appendix B), and hence we interpret them as Stratonovich equations which
allow us to apply ordinary calculus.

In order to better visualize the free parameters of the model, let us define
new rescaled time and amplitudes through%
\begin{equation}
\tau=\gamma_{\mathrm{s}}t\text{, }\beta_{\pm1}=\frac{\chi}{\sqrt
{\gamma_{\mathrm{p}}\gamma_{\mathrm{s}}}}\alpha_{\pm1}\text{, }\beta_{0}%
=\frac{\chi}{\gamma_{\mathrm{s}}}\alpha_{0}\text{,} \label{Rescaled}%
\end{equation}
and similar expressions for the conjugate fields $\alpha_{\pm1}^{+}$ and
$\alpha_{0}^{+}$. Note that as time is measured in units of $\gamma
_{\mathrm{s}}^{-1}$, the noises must be rescaled also as%
\begin{equation}
\zeta_{\pm1}\left(  \tau\right)  =\frac{1}{\sqrt{\gamma_{\mathrm{s}}}}\xi
_{\pm1}\left(  t\right)  ,
\end{equation}
and similarly for $\xi_{\pm1}^{+}$, if we want to preserve their statistical
properties (\ref{NoiseCorr}) in terms of the adimensional time $\tau$.

Finally, these changes make Langevin equations (\ref{Langevin}) read (of
course, derivatives are respect with the new dimensionless time)%
\begin{align}
\dot{\beta}_{0}  &  =\kappa\left(  \sigma-\beta_{0}-\beta_{+1}\beta
_{-1}\right) \label{ReescaledLangevin}\\
\dot{\beta}_{0}^{+}  &  =\kappa\left(  \sigma-\beta_{0}^{+}-\beta_{+1}%
^{+}\beta_{-1}^{+}\right) \nonumber\\
\dot{\beta}_{+1}  &  =-\beta_{+1}+\beta_{0}\beta_{-1}^{+}+g\sqrt{\beta_{0}%
}\zeta\left(  t\right) \nonumber\\
\dot{\beta}_{+1}^{+}  &  =-\beta_{+1}^{+}+\beta_{0}^{+}\beta_{-1}+g\sqrt
{\beta_{0}}\zeta^{+}\left(  t\right) \nonumber\\
\dot{\beta}_{-1}  &  =-\beta_{-1}+\beta_{0}\beta_{+1}^{+}+g\sqrt{\beta_{0}%
}\zeta^{\ast}\left(  t\right) \nonumber\\
\dot{\beta}_{-1}^{+}  &  =-\beta_{-1}^{+}+\beta_{0}^{+}\beta_{+1}+g\sqrt
{\beta_{0}}\left[  \zeta^{+}\left(  t\right)  \right]  ^{\ast},\nonumber
\end{align}
which have only the following 3 dimensionless parameters:%
\begin{equation}
\kappa=\frac{\gamma_{\mathrm{p}}}{\gamma_{\mathrm{s}}}\text{, }\sigma
=\frac{\mathcal{E}_{\mathrm{p}}\chi}{\gamma_{\mathrm{p}}\gamma_{\mathrm{s}}%
}\text{ and }g=\frac{\chi}{\sqrt{\gamma_{\mathrm{p}}\gamma_{\mathrm{s}}}%
}\text{.}%
\end{equation}

\section{Classical emission: rotational symmetry breaking}

Before analyzing the quantum properties of the system, let us examine the
emission of this DOPO as predicted by classical optics. It is possible to
retrieve the classical equations of the 2-transverse-mode DOPO from the
quantum Langevin ones (\ref{ReescaledLangevin}) by setting all the noises to
zero and making $\beta_{m}^{+}=\beta_{m}^{\ast}$ ($m$ refers to any mode of
the problem). We are interested in finding the stationary emission, and then
set to zero the time derivatives of the amplitudes, arriving to%
\begin{align}
\beta_{0} &  =\sigma-\beta_{+1}\beta_{-1}\label{ClassicalEqs}\\
\beta_{+1} &  =\beta_{0}\beta_{-1}^{\ast}\nonumber\\
\beta_{-1} &  =\beta_{0}\beta_{+1}^{\ast}.\nonumber
\end{align}

Before solving these equations, let us stress that these are invariant under
the following transformation%
\begin{equation}
\beta_{\pm1}\longrightarrow e^{\pm i\psi}\beta_{\pm1},
\end{equation}
and hence the phase difference between opposite OAM modes is classically undetermined.

By decomposing into modulus and phase the amplitudes $\beta_{m}$, it is
straightforward to find the solutions of the system (\ref{ClassicalEqs}). If
in addition a linear stability analysis is made, it is easy to find the
following result:

\begin{itemize}
\item For $\sigma<1$ the only stable solution is%
\begin{align}
\bar{\beta}_{0}  &  =\sigma\\
\bar{\beta}_{\pm1}  &  =0,\nonumber
\end{align}
and hence the signal modes are off if\ the pump $\mathcal{E}_{\mathrm{p}}$
doesn't exceed a threshold value $\mathcal{E}_{th}=\gamma_{\mathrm{p}}%
\gamma_{\mathrm{s}}/\chi$.

\item On the other hand, if $\sigma>1$, the signal modes are switched on, and
the only stable solution in this case is%
\begin{align}
\bar{\beta}_{0}  &  =1\label{AboveThreshold}\\
\bar{\beta}_{\pm1}  &  =\rho e^{\mp i\theta}\text{ with }\rho=\sqrt{\sigma
-1},\nonumber
\end{align}
where, as stated above, $\theta$ is arbitrary.
\end{itemize}

If we substitute the last result for the signal modes in the expression of the
corresponding slowly varying envelope (\ref{SVEs}), we find the generated
pattern to be%
\begin{equation}
\bar{A}_{\mathrm{s}}\left(  \mathbf{r}\right)  =\rho\left[  e^{-i\theta}%
L_{+1}\left(  \mathbf{r}\right)  +e^{+i\theta}L_{-1}\left(  \mathbf{r}\right)
\right]  =\sqrt{2}\rho H_{10}^{\theta}\left(  \mathbf{r}\right)  \text{,}%
\end{equation}
which is a TEM$_{10}$ mode forming an angle $\theta$ with respect to the $x$
axis (\ref{HG}) as shown in Fig. 1. Due to the arbitrariness of $\theta$, the
pattern can arise with any orientation; this reflects the rotational
invariance of the system, wich in turn is broken after the TEM$_{10}$ mode generation.

In the following, we will call $H_{10}^{\theta}\left(  \mathbf{r}\right)  $
the \textit{bright mode} (as it is classically excited) and its orthogonal
mode $H_{01}^{\theta}\left(  \mathbf{r}\right)  $ the \textit{dark mode} (as
it is classically empty of photons), and will define the collective indices
$\mathrm{b}=\left(  10,\theta\right)  $ and $\mathrm{d}=\left(  01,\theta
\right)  $ to simplify the notation.

\section{Quantum properties: pattern diffusion and non-critical squeezing}

\subsection{The linearized Langevin equations}

We are going to discuss the quantum properties of the down converted field by
inspection of the quantum Langevin equations (\ref{ReescaledLangevin}) in the
limit $\gamma_{\mathrm{p}}\gg\gamma_{\mathrm{s}}$, i.e., $\kappa\gg1$, where
the pump variables can be adiabatically eliminated. We will show later
(Section V) that all the important properties found in this limit are valid in general.

One common way of treating the adiabatic elimination of the pump consists on
setting $\dot{\beta}_{0}=\dot{\beta}_{0}^{+}=0$ in the first two equations of
(\ref{ReescaledLangevin}). However, this method has a problem: The initial
Langevin equations are equal in either Ito or Stratonovich forms, but the new
Langevin equations obtained for the signal modes doesn't have this property
(see Appendix B). Hence it is fair to ask within which interpretation (Ito or
Stratonovich) is that procedure correct, if it is correct at all. As proved in
\cite{German09} by performing the adiabatic elimination in the Fokker-Planck
equation (where there are no problems of interpretation), the usual method for
eliminating the pump is correct within Ito's interpretation, and hence the
Stratonovich form of the stochastic equations satisfied by the signal modes
only has an unexpected extra term proportional to $g^{2}$ (see Appendix B)
and reads:%
\begin{align}
\dot{\beta}_{+1}  &  =-\left(  1-g^{2}/4\right)  \beta_{+1}+\beta_{0}%
\beta_{-1}^{+}+g\sqrt{\beta_{0}}\zeta\left(  \tau\right) \label{AdElEqs}\\
\dot{\beta}_{+1}^{+}  &  =-\left(  1-g^{2}/4\right)  \beta_{+1}^{+}+\beta
_{0}^{+}\beta_{-1}+g\sqrt{\beta_{0}^{+}}\zeta^{+}\left(  \tau\right)
\nonumber\\
\dot{\beta}_{-1}  &  =-\left(  1-g^{2}/4\right)  \beta_{-1}+\beta_{0}%
\beta_{+1}^{+}+g\sqrt{\beta_{0}}\zeta^{\ast}\left(  \tau\right) \nonumber\\
\dot{\beta}_{-1}^{+}  &  =-\left(  1-g^{2}/4\right)  \beta_{-1}^{+}+\beta
_{0}^{+}\beta_{+1}+g\sqrt{\beta_{0}^{+}}\left[  \zeta^{+}\left(  \tau\right)
\right]  ^{\ast},\nonumber
\end{align}
with%
\begin{align}
\beta_{0}  &  =\sigma-\beta_{+1}\beta_{+1},\\
\beta_{0}^{+}  &  =\sigma-\beta_{+1}^{+}\beta_{+1}^{+}.\nonumber
\end{align}

In order to find analytic predictions from these equations, we assume well
above threshold operating conditions, where classical emission dominates, thus
allowing for a linearization procedure. The correct way to perform this was
described in \cite{NonlinearDOPO,NonlinearOPO}, and relies on the smallness of
the $g$ parameter (see Appendix A). In particular, making a perturbative
expansion of the amplitudes in terms of this parameter, it is possible to show
that the classical equations are recovered as the $g^{0}$ order of the
expansion in Eqs. (\ref{AdElEqs}), while the linear quantum correction appears
in the $g^{1}$ order (note that the noise term is already of order $g^{1}$).

Hence, the usual linearization procedure begins by writing the amplitudes as
$\beta_{m}=\bar{\beta}_{m}+\delta\beta_{m}$ and $\beta_{m}^{+}=\bar{\beta}%
_{m}^{\ast}+\delta\beta_{m}^{+}$, and treat the fluctuations as order $g$
perturbations. However, as stated in the Introduction, in the system we are
dealing with we expect that quantum noise rotates the generated TEM$_{10}$
mode, and hence fluctuations of the fields in an arbitrary direction of phase
space could not be small (i.e., order $g$). Nevertheless, Eqs. (\ref{AdElEqs})
can be linearized if the amplitudes are expanded as%
\begin{align}
\beta_{\pm1}\left(  \tau\right)   &  =\left[  \rho+b_{\pm1}\left(
\tau\right)  \right]  e^{\mp i\theta\left(  \tau\right)  }\label{LinExp}\\
\beta_{\pm1}^{+}\left(  \tau\right)   &  =\left[  \rho+b_{\pm1}^{+}\left(
\tau\right)  \right]  e^{\pm i\theta\left(  \tau\right)  },\nonumber
\end{align}
because as we will prove $\theta\left(  \tau\right)  $ carries the larger part
of the fluctuations, while the $b$'s and $\dot{\theta}$ remain as order $g$
quantities. In addition, expanding the fields in this way allows us to track
the evolution of the classical pattern's orientation, as we take $\theta$ as
an explicit quantum variable. Then, writing Eqs. (\ref{AdElEqs}) up to order
$g$, we arrive to the following linear system (arriving to this expression is
not as straightforward as it might seem, see Appendix C)%
\begin{equation}
-2i\rho\mathbf{w}_{0}\dot{\theta}+\mathbf{\dot{b}=}\mathcal{L}\mathbf{b}%
+g\boldsymbol{\zeta}\left(  \tau\right)  ,\label{LinLan}%
\end{equation}
with%
\begin{equation}
\mathbf{b}=%
\begin{pmatrix}
b_{+1}\\
b_{+1}^{+}\\
b_{-1}\\
b_{-1}^{+}%
\end{pmatrix}
\text{\ , }\boldsymbol{\zeta}\left(  \tau\right)  =%
\begin{pmatrix}
\zeta\left(  \tau\right)  \\
\zeta^{+}\left(  \tau\right)  \\
\zeta^{\ast}\left(  \tau\right)  \\
\left[  \zeta^{+}\left(  \tau\right)  \right]  ^{\ast}%
\end{pmatrix}
,
\end{equation}
and where $\mathcal{L}$ is a real, symmetric matrix given by%
\begin{equation}
\mathcal{L}=-%
\begin{pmatrix}
\sigma & 0 & \sigma-1 & -1\\
0 & \sigma & -1 & \sigma-1\\
\sigma-1 & -1 & \sigma & 0\\
-1 & \sigma-1 & 0 & \sigma
\end{pmatrix},
\end{equation}
with the following eigensystem%
\begin{equation}%
\begin{array}
[c]{ll}%
\lambda_{0}=0, & \mathbf{w}_{0}=\frac{1}{2}\operatorname{col}\left(
1,-1,-1,1\right)  \\
\lambda_{1}=-2, & \mathbf{w}_{1}=\frac{1}{2}\operatorname{col}\left(
1,1,-1,-1\right)  \\
\lambda_{2}=-2\left(  \sigma-1\right)  , & \mathbf{w}_{2}=\frac{1}%
{2}\operatorname{col}\left(  1,1,1,1\right)  \\
\lambda_{3}=-2\sigma, & \mathbf{w}_{3}=\frac{1}{2}\operatorname{col}\left(
1,-1,1,-1\right)  .
\end{array}
\label{Eigensystem}%
\end{equation}

Defining the projections $c_{m}\left(  \tau\right)  =\mathbf{w}_{m}%
\cdot\mathbf{b}\left(  \tau\right)  $, and projecting the linear system
(\ref{LinLan}) onto these eigenmodes, we find the following set of decoupled
linear equations ($c_{0}$ is set to zero, as otherwise it would just entail a
redefinition of $\theta$)
\begin{subequations}
\label{ProjLinLan}%
\begin{align}
\dot{\theta}  &  \mathbf{=}\frac{g}{2\rho}\eta_{0}\left(  \tau\right)
\label{ThetaEvo}\\
\dot{c}_{1}  &  =-2c_{1}+ig\eta_{1}\left(  \tau\right) \label{c1evo}\\
\dot{c}_{2}  &  =-2\left(  \sigma-1\right)  c_{2}+g\eta_{2}\left(  \tau\right)
\label{c2evo}\\
\dot{c}_{3}  &  =-2\sigma c_{3}+g\eta_{3}\left(  \tau\right),  \label{c3evo}%
\end{align}
where the following real noises have been defined%
\end{subequations}
\begin{align}
\eta_{0}\left(  \tau\right)   &  =i\mathbf{w}_{0}\cdot\boldsymbol{\zeta
}\left(  \tau\right)  =\operatorname{Im}\left\{  \zeta^{+}\left(  \tau\right)
-\zeta\left(  \tau\right)  \right\} \\
\eta_{1}\left(  \tau\right)   &  =-i\mathbf{w}_{1}\cdot\boldsymbol{\zeta
}\left(  \tau\right)  =\operatorname{Im}\left\{  \zeta^{+}\left(  \tau\right)
+\zeta\left(  \tau\right)  \right\} \nonumber\\
\eta_{2}\left(  \tau\right)   &  =\mathbf{w}_{2}\cdot\boldsymbol{\zeta}\left(
\tau\right)  =\operatorname{Re}\left\{  \zeta\left(  \tau\right)  +\zeta
^{+}\left(  \tau\right)  \right\} \nonumber\\
\eta_{3}\left(  \tau\right)   &  =\mathbf{w}_{3}\cdot\boldsymbol{\zeta}\left(
\tau\right)  =\operatorname{Re}\left\{  \zeta\left(  \tau\right)  -\zeta
^{+}\left(  \tau\right)  \right\}  ,\nonumber
\end{align}
which have zero mean and nonzero correlations%
\begin{equation}
\left\langle \eta_{m}\left(  \tau_{1}\right)  \eta_{n}\left(  \tau_{2}\right)
\right\rangle =\delta_{mn}\delta\left(  \tau_{1}-\tau_{2}\right)  .
\label{RealNoiseStat}%
\end{equation}

Note finally that the solutions for the projections $c_{j}$ will be of order
$g$ (see Appendix D), and hence so will be the $b$'s (note that this is not
the case for $\theta$, whose initial value is completely arbitrary, although
its variation $\dot{\theta}$ is indeed of order $g$). This is consistent with
the initial assumptions about the orders in $g$ of the involved quantities.

\subsection{Quantum diffusion of the classical pattern}

The first quantum effect that we are going to show concerns the orientation of
the classically generated mode. Equation (\ref{ThetaEvo}) defines a
\textit{Wiener process} for $\theta$, \ thus showing that the orientation of
this bright mode diffuses with time ruled by quantum noise.

How fast this diffusion is can be measured by evaluating the variance of
$\theta$. Using the statistical properties of noise (\ref{RealNoiseStat}), it
is straightforward to obtain the following result%
\begin{equation}
V_{\theta}\left(  \tau\right)  =\left\langle \delta\theta^{2}\left(
\tau\right)  \right\rangle =D\tau\text{,}\label{Theta Var}%
\end{equation}
where $D=d/\left(  \sigma-1\right)  $ with
\begin{equation}
d=g^{2}/4=\chi^{2}/4\gamma_{\mathrm{p}}\gamma_{\mathrm{s}}.\label{d}%
\end{equation}
In (\ref{Theta Var}) the notation $\delta A=A-\left\langle A\right\rangle $
has been used.

Eq. (\ref{Theta Var}) shows that the delocalization of the pattern's orientation
increases as time passes by, though for typical system parameters (see
Appendix A) one finds $d\simeq4\cdot10^{-13}$, and hence the rotation of the
pattern will be fast only when working terribly close to threshold
\cite{CompDifPRL}.

\subsection{Squeezing properties of the 2-transverse-mode DOPO}

\textbf{Definition and criterium for squeezing.}\emph{ }In the Introduction we
defined squeezed light as that having some quadrature fluctuating below the
vacuum level. Hence, in order to find out whether a light beam is in a
squeezed state or not, one has to measure its \textit{quadrature}
\textit{fluctuations} and then compare them with the value set by the vacuum
state. The question then is: What quantity accounts for quadrature
fluctuations of a light beam? Of course, in the case of a single mode of light
this quantity can be directly the uncertainty of the quadrature. However,
outside the cavity there exists a continuum of modes, and the quantity
accounting for these fluctuations has to be adapted to what can be most easily
observed in an experiment.

As was first shown in \cite{Collett84}, the quantity accounting for
fluctuations of quadrature $\hat{X}_{m}^{\varphi}$ outside the cavity ($m$
refers to any signal spatial mode of our 2-transverse-mode DOPO) is%
\begin{equation}
V^{\mathrm{out}}\left(  \omega;\hat{X}_{m}^{\varphi}\right)  =1+S_{m}%
^{\varphi}\left(  \omega\right)  ,\label{Vout}%
\end{equation}
with
\begin{equation}
S_{m}^{\varphi}\left(  \omega\right)  =\frac{2}{g^{2}}\int_{-\infty}^{+\infty
}d\bar{\tau}\left\langle :\delta\hat{X}_{m}^{\varphi}\left(  \tau\right)
\delta\hat{X}_{m}^{\varphi}\left(  \tau+\bar{\tau}\right)  :\right\rangle
e^{-i\omega\bar{\tau}}\label{Susual}%
\end{equation}
where the factor $g^{-2}$ appears after including our rescaled variables
(\ref{Rescaled}).

We will call $V^{\mathrm{out}}$ the \textit{noise spectrum} and $S_{m}%
^{\varphi}\left(  \omega\right)  $ the \textit{squeezing spectrum}. Frequency
$\omega$ is usually called \textit{noise frequency}, and it must not be
confused with the optical frequency. In fact, noise frequency $\omega$ has
contributions of every pair of modes lying in opposite sidebands around the
optical frequency $\omega_{0}+\omega$, where $\omega_{0}$ is the carrier
frequency of the detected beam (don't miss that in our case this frequency is
measured in units of $\gamma_{\mathrm{s}}$).

$V^{\mathrm{out}}$ can be measured via a balanced homodyne detection
experiment: A coherent, intense field (local oscillator field) prepared in
mode $m$, and with a phase $\varphi$ is mixed with the beam exiting the DOPO
in a 50/50 beam splitter; it is easy to show that the operator associated to
the intensity difference between the two output ports of the beam splitter is
proportional to the quadrature $\hat{X}_{m}^{\varphi,\mathrm{out}}$ of the
beam exiting the DOPO. Hence, the normalized correlation spectrum of the
intensity difference (which can be measured with a simple spectrum analyzer)
coincides with expression (\ref{Vout}). Factor 2 in (\ref{Susual}) comes from
the relation between the intracavity and the output modes when the input modes
(the pump laser and the outer modes) are coherent; this factor is
$2\gamma_{\mathrm{s}}$ when time (frequency) is not normalized to
$\gamma_{\mathrm{s}}^{-1}$($\gamma_{\mathrm{s}}$).

In the above expression for the squeezing spectrum, it is assumed that the
emission of the DOPO reaches a stationary state (i.e., the correlation
function of $\delta\hat{X}_{m}^{\varphi}$ at two different times $\tau$ and
$\tau+\bar{\tau}$ depends only on the time difference $\left\vert \bar{\tau
}\right\vert $) and that the observation time, say $T$, is large as compared
with the coherence time of $\delta\hat{X}_{m}^{\varphi}$. In the
2-transverse-mode DOPO there exists an undamped quantity excited by quantum
noise, $\theta$, and hence, any quantity with explicit dependence on it will
not arrive to a stationary state. Thus, under some circumstances we will need
the expression of the squeezing spectrum measured when the field is not
stationary, which was derived in \cite{Gea-Banacloche90}. In that reference it
is shown that when the measurement extends during a time interval $T$
(normalized to $\gamma_{\mathrm{s}}^{-1}$ in our expressions) the squeezing
spectrum reads%
\begin{align}
S_{m}^{\varphi}\left(  \omega\right)  =\frac{2}{g^{2}T}\int_{0}^{T}\int
_{0}^{T}d\tau d\tau^{\prime} &  \left\langle :\delta\hat{X}_{m}^{\varphi
}\left(  \tau\right)  \delta\hat{X}_{m}^{\varphi}\left(  \tau^{\prime}\right)
:\right\rangle \nonumber\\
&  \times\cos\left[  \omega\left(  \tau-\tau^{\prime}\right)  \right]
,\label{Sgen}%
\end{align}
which obviously reduces to (\ref{Susual}) if the emission is stationary.

For the vacuum state $V^{\mathrm{out}}=1$, hence if $V^{\mathrm{out}}\left(
\tilde{\omega};\hat{X}_{m}^{\varphi}\right)  <1$ for some frequency
$\tilde{\omega}$, we can say that the detected light is in a squeezed state
for mode $m$ at noise frequency $\tilde{\omega}$.

Finally, let us stress that we will evaluate quantum correlations as those
involved in (\ref{Sgen}) via stochastic correlations, as allowed by property
(\ref{N-St}). Hence, in the following we will change operators by their
equivalent stochastic variables within the positive \textit{P} representation
in all the definitions above, and their associated normally ordered
correlations by the corresponding stochastic correlations.

\textbf{Independent quadratures and non critical squeezing. }A first step
towards analyzing the squeezing properties of the field within the framework
presented above, is identifying a set of independent quadratures. This is easy
in our case, as the eigenmodes $\mathbf{w}_{m}$ give us a set of quadratures
with well defined squeezing properties. In particular, from (\ref{LinExp}),
(\ref{LGtoHG}), and (\ref{Quadrature}), it is easy to find the following
relations%
\begin{align}
X_{\mathrm{b}}\left(  \tau\right)   &  =2\sqrt{2}\rho+\sqrt{2}c_{2}\left(
\tau\right) \\
Y_{\mathrm{b}}\left(  \tau\right)   &  =-i\sqrt{2}c_{3}\left(  \tau\right)
\nonumber\\
X_{\mathrm{d}}\left(  \tau\right)   &  =i\sqrt{2}c_{0}\left(  \tau\right)
\nonumber\\
Y_{\mathrm{d}}\left(  \tau\right)   &  =\sqrt{2}c_{1}\left(  \tau\right)
,\nonumber
\end{align}
where $\left(  X_{\mathrm{b}},Y_{\mathrm{b}}\right)  $ and $\left(
X_{\mathrm{d}},Y_{\mathrm{d}}\right)  $ are the amplitude and phase
quadratures of the bright and dark modes, $H_{10}^{\theta}\left(
\mathbf{r}\right)  $ and $H_{01}^{\theta}\left(  \mathbf{r}\right)  $, respectively.

The evolution of these quadratures can be found from the equations satisfied
by the projections (\ref{ProjLinLan}), which are solved in Appendix D
(remember that $c_{0}=0$). In particular, there we show that after some time
these projections reach a stationary state, and hence we can evaluate the
noise spectrum of the quadratures by using the stationary expression for the
squeezing spectrum (\ref{Susual}). Using the results found in Appendix D for
the correlation spectrum of the projections (\ref{CSpectrum}), it is
straightforward to find the following \ results
\begin{subequations}
\label{NoiseSpectra}%
\begin{align}
V^{\mathrm{out}}\left(  \omega;X_{\mathrm{b}}\right)   &  =1+\frac{1}{\left(
\sigma-1\right)  ^{2}+\omega^{2}/4}\label{XcSpectrum}\\
V^{\mathrm{out}}\left(  \omega;Y_{\mathrm{b}}\right)   &  =1-\frac{1}%
{\sigma^{2}+\omega^{2}/4}\label{YcSpectrum}\\
V^{\mathrm{out}}\left(  \omega;X_{\mathrm{d}}\right)   &  =1
\label{XsSpectrum}\\
V^{\mathrm{out}}\left(  \omega;Y_{\mathrm{d}}\right)   &  =1-\frac{1}%
{1+\omega^{2}/4}. \label{YsSpectrum}%
\end{align}

We see that the quadratures of the bright mode have the same behavior as
those of the single mode DOPO: the phase quadrature $Y_{\mathrm{b}}$ is
perfectly squeezed $\left(  V^{\mathrm{out}}=0\right)  $ at zero noise
frequency $\left(  \omega=0\right)  $ only at the bifurcation $\left(
\sigma=1\right)  $.

On the other hand, the dark mode has perfect squeezing in its phase quadrature
$Y_{\mathrm{d}}$ at zero noise frequency. What is interesting is that this
result is independent of the distance from threshold, and thus, it is a
non-critical phenomenon.

This result was first shown in \cite{PRL}, and we can understand it by
following the reasoning given in the Introduction: As the orientation $\theta$
of the classically excited pattern is undetermined in the long-time limit, its
OAM must be fully determined at low noise frequencies. On the other hand, the
OAM of the bright mode $H_{10}^{\theta}\left(  \mathbf{r}\right)  $ is nothing
but its $\pi/2$ phase shifted orthogonal HG mode, i.e., $-i\partial_{\phi}%
H_{10}^{\theta}\left(  \mathbf{r}\right)  =iH_{01}^{\theta}\left(
\mathbf{r}\right)  $, which used as a local oscillator in an homodyne
detection experiment would lead to the observation of the
$Y_{\mathrm{d}}$ quadrature fluctuations.

\textbf{On canonical pairs and noise transfer}. Although all the results we
have discussed up to now were expected to occur after the arguments given in
the Introduction, a strange unexpected result has appeared in
(\ref{NoiseSpectra}): The quadratures of the dark mode seem to violate the
uncertainty principle as $V^{\mathrm{out}}\left(  \omega=0;X_{\mathrm{d}%
}\right)  \cdot V^{\mathrm{out}}\left(  \omega=0;Y_{\mathrm{d}}\right)  =0$.

Moreover, the noise spectrum of an arbitrary quadrature of the dark mode,
which can be written in terms of its amplitude and phase quadratures as
$X_{\mathrm{d}}^{\varphi}=X_{\mathrm{d}}\cos\varphi+Y_{\mathrm{d}}\sin\varphi
$, is%
\end{subequations}
\begin{equation}
V^{\mathrm{out}}\left(  \omega;X_{\mathrm{d}}^{\varphi}\right)  =1-\frac
{\sin^{2}\varphi}{1+\omega^{2}/4}\text{,}%
\end{equation}
what shows that all the quadratures of the dark mode (except its amplitude
quadrature) are squeezed. Thus two orthogonal quadratures cannot form a
canonical pair as they satisfy the relation
\begin{equation}
V^{\mathrm{out}}\left(  \omega;X_{\mathrm{d}}^{\varphi}\right)  \cdot
V^{\mathrm{out}}\left(  \omega;X_{\mathrm{d}}^{\varphi+\pi/2}\right)  <1,
\end{equation}
in clear violation of the uncertainty principle.

The natural question now is: Where does the excess of noise go if it is not
transferred from one quadrature to its orthogonal one? The answer is that it
goes to the pattern orientation, which is actually fully undetermined in the
long term as we showed above (\ref{Theta Var}). This section is devoted to
prove this statement.

In particular we will prove that two orthogonal quadratures of the dark mode
do not form a canonical pair, while the orientation $\theta$ is the canonical
pair of all the squeezed quadratures. One way to prove this would be to
evaluate the commutator between the quantum operators involved. However,
$\theta$ is half the phase difference between the opposite OAM modes
$\beta_{\pm1}$, whose associated operator has a very difficult expression
\cite{Luis93Yu97}, making the calculation of the needed commutators quite
hard. Nevertheless, in \cite{arXiv} we developed a simple proof based on
classical field methods, and this is the one we present here.

A usual approach one uses to move from classical to quantum optics is to
change the classical normal variables for each mode of the field, $\beta_{m}$
and $\beta_{m}^{\ast}$, by boson operators $\hat{a}_{m}$ and $\hat{a}%
_{m}^{\dagger}$, satisfying commutation relations%
\begin{equation}
\left[  \hat{a}_{m},\hat{a}_{m}^{\dagger}\right]  =i\left\{  \beta_{m}%
,\beta_{m}^{\ast}\right\}  ,
\end{equation}
where $\left\{  f,h\right\}  $ denotes the \textit{Poisson bracket} between
two functions $f\left(  \beta_{m},\beta_{m}^{\ast}\right)  $ and $h\left(
\beta_{m},\beta_{m}^{\ast}\right)  $ defined as%
\begin{equation}
\left\{  f,h\right\}  =\frac{1}{i}%
{\displaystyle\sum\limits_{m}}
\frac{\partial f}{\partial\beta_{m}}\frac{\partial h}{\partial\beta_{m}^{\ast
}}-\frac{\partial f}{\partial\beta_{m}^{\ast}}\frac{\partial h}{\partial
\beta_{m}}.
\end{equation}

As an example, the Poisson bracket of two monomode orthogonal quadratures
$X_{m}=\beta_{m}+\beta_{m}^{\ast}$ and $Y_{m}=-i\left(  \beta_{m}-\beta
_{m}^{\ast}\right)  $ is found to be%
\begin{equation}
\left\{  X_{m},Y_{m}\right\}  =2.
\end{equation}
In general two phase space functions $f\left(  \beta_{m},\beta_{m}^{\ast
}\right)  $ and $h\left(  \beta_{m},\beta_{m}^{\ast}\right)  $ are said to
form a canonical pair if their Poisson bracket is of the form%
\begin{equation}
\left\{  f,h\right\}  =C, \label{CanBraRel}%
\end{equation}
with $C$ a real number.

In our case, the functions we are interested in are the classical counterparts
of the dark mode quadratures, which using (\ref{LGtoHGs}) with $\psi=\theta$
are written as
\begin{equation}
X_{\mathrm{d}}^{\varphi}=\frac{i}{\sqrt{2}}\left[  e^{-i\varphi}\left(
e^{i\theta}\beta_{+1}-e^{-i\theta}\beta_{-1}\right)  \right]  +\mathrm{c.c.},
\end{equation}
with the orientation given by%
\begin{equation}
\theta=\frac{1}{2i}\ln\left[  \frac{\beta_{+1}^{\ast}\beta_{-1}}{\sqrt
{\beta_{+1}^{\ast}\beta_{+1}}\sqrt{\beta_{-1}^{\ast}\beta_{-1}}}\right]  .
\end{equation}

Now using the definition of the Poisson brackets with $m=\pm1$ and after some
algebra, it is possible to show that%
\begin{equation}
\left\{  X_{\mathrm{d}}^{\varphi},X_{\mathrm{d}}^{\varphi+\pi/2}\right\}
=\frac{\left\vert \beta_{-1}\right\vert -\left\vert \beta_{+1}\right\vert
}{2\left\vert \beta_{+1}\right\vert \left\vert \beta_{-1}\right\vert },
\end{equation}
and%
\begin{equation}
\left\{  X_{\mathrm{d}}^{\varphi},\theta\right\}  =\frac{i\left(  \left\vert
\beta_{+1}\right\vert +\left\vert \beta_{-1}\right\vert \right)  \left(
e^{i\varphi}\left\vert \beta_{-1}\right\vert -e^{-i\varphi}\left\vert
\beta_{-1}\right\vert \right)  }{4\sqrt{2}\left\vert \beta_{+1}\right\vert
^{3/2}\left\vert \beta_{-1}\right\vert ^{3/2}},
\end{equation}
with $\left\vert \beta_{m}\right\vert =\sqrt{\beta_{m}^{\ast}\beta_{m}}$.

On the other hand, in the two-transverse-mode DOPO, the number of photons with
opposite OAM is sensibly equal, i.e., $\left\vert \beta_{+1}\right\vert
\approx\left\vert \beta_{-1}\right\vert $; hence, the dominant term of the
previous brackets will be that with $\left\vert \beta_{+1}\right\vert
=\left\vert \beta_{-1}\right\vert =\rho$, and thus%
\begin{equation}
\left\{  X_{\mathrm{d}}^{\varphi},X_{\mathrm{d}}^{\varphi+\pi/2}\right\}
\approx0
\end{equation}
and%
\begin{equation}
\left\{  X_{\mathrm{d}}^{\varphi},\theta\right\}  \approx-\frac{\sin\varphi
}{\sqrt{2}\rho}\text{.}%
\end{equation}

Comparing these with (\ref{CanBraRel}) we find that this is a confirmation of
what we expected: two orthogonal dark quadratures $X_{\mathrm{d}}^{\varphi}$
and $X_{\mathrm{d}}^{\varphi+\pi/2}$ do not form a canonical pair (moreover,
they commute), while $X_{\mathrm{d}}^{\varphi}$ and $\theta$ do. This can be
seen as an indirect proof of the same conclusion for the corresponding operators.

\textbf{Homodyne detection with a fixed local oscillator}. So far we have
considered the situation in which one is able to detect independently the
bright and dark modes. However, as shown by Eq. (\ref{ThetaEvo}), these modes
are rotating randomly, what means that a local oscillator field following that
random rotation should be used in order to detect them separately. This might
be a really complicated, if not impossible, task, so we analyze now the more
realistic situation in which the local oscillator is matched to the orthogonal
orientation of the emerging pattern only at some initial time, remaining then
with the same orientation during the observation time. We will show that even
in this case, and as the rotation of the modes is quite slow (\ref{Theta Var}%
), large levels of noise reduction can be obtained.

Without loss of generality, we suppose that the bright mode emerges from the
resonator at some initial time $\tau=0$ oriented within the $x$ axis, i.e.,
$\theta\left(  0\right)  =0$. Hence, by using a TEM$_{01}$ local oscillator
with a phase $\varphi$, the quadrature $X_{01}^{\varphi}$ of a fixed
$H_{01}\left(  \mathbf{r}\right)  $ mode (the initially dark mode) will be
measured. In terms of the Gauss-Laguerre modes, the amplitude $\beta_{01}$ of
the this mode is given by (\ref{LGtoHGs}) with $\psi=0$. Then, by using the
expansion (\ref{LinExp}) of the amplitudes $\beta_{m}$ as functions of
the fluctuations $b_{m}$ and the orientation $\theta$, quadrature $X_{01}^{\varphi}$
of this mode can be rewritten as%
\begin{align}
X_{01}^{\varphi}  &  =2\sqrt{2}\rho\cos\varphi\sin\theta+\sqrt{2}c_{2}%
\cos\varphi\sin\theta\\
&  +\sqrt{2}c_{1}\sin\varphi\cos\theta-i\sqrt{2}c_{3}\sin\varphi\sin
\theta\text{.}\nonumber
\end{align}
Hence, the 2-time correlation function of $X_{01}^{\varphi}$ yields (note that
(\ref{ProjLinLan}) shows that $c_{1}$, $c_{2}$, $c_{3}$, $\sin\theta$ and
$\cos\theta$ are uncorrelated)%
\begin{align}
&  \left\langle X_{01}^{\varphi}\left(  \tau_{1}\right)  X_{01}^{\varphi
}\left(  \tau_{2}\right)  \right\rangle =2\cos^{2}\varphi S\left(  \tau
_{1},\tau_{2}\right)  \left[  4\rho^{2}+C_{2}\left(  \tau_{1},\tau_{2}\right)
\right] \label{StillCorrelation}\\
&  \text{ \ \ }+2\sin^{2}\varphi\left[  C_{1}\left(  \tau_{1},\tau_{2}\right)
C\left(  \tau_{1},\tau_{2}\right)  -C_{3}\left(  \tau_{1},\tau_{2}\right)
S\left(  \tau_{1},\tau_{2}\right)  \right]  ,\nonumber
\end{align}
with%
\begin{align}
S\left(  \tau_{1},\tau_{2}\right)   &  =\left\langle \sin\theta\left(
\tau_{1}\right)  \sin\theta\left(  \tau_{2}\right)  \right\rangle
\label{StillCorr}\\
C\left(  \tau_{1},\tau_{2}\right)   &  =\left\langle \cos\theta\left(
\tau_{1}\right)  \cos\theta\left(  \tau_{2}\right)  \right\rangle \nonumber\\
C_{m}\left(  \tau_{1},\tau_{2}\right)   &  =\left\langle c_{m}\left(  \tau
_{1}\right)  c_{m}\left(  \tau_{2}\right)  \right\rangle .\nonumber
\end{align}

As shown in Appendices D and E, the latter correlation functions can be
evaluated by using the linear evolution equations of the projections $c_{m}$
and $\theta$ -see Eqs. (\ref{Ccorr}), (\ref{SinCorr}) and (\ref{CosCorr})-;
then from (\ref{StillCorrelation}) the squeezing spectrum of $X_{01}^{\varphi
}$ can be found by using the general expression (\ref{Sgen}), as in this case
the stationary expression (\ref{Susual}) cannot be used because $S\left(
\tau_{1},\tau_{2}\right)  $ and $C\left(  \tau_{1},\tau_{2}\right)  $ don't
reach a stationary state.

Although the general expression for $S_{01}^{\varphi}\left(  \omega\right)  $
is too lengthy to be written here, a more compact approximated expression can
be found in the limit of small $d$, leading to the following expression for
the noise spectrum%
\begin{equation}
V^{\mathrm{out}}\left(  \omega;X_{01}^{\varphi}\right)  =1+S_{01}^{0}\left(
\omega\right)  \cos^{2}\varphi+S_{01}^{\pi/2}\left(  \omega\right)  \sin
^{2}\varphi, \label{FixedNoiseSpectrum}%
\end{equation}
with%
\begin{equation}
S_{01}^{0}=\frac{8}{\omega^{2}}\left(  1-\operatorname{sinc}\omega T\right)
-\frac{4dT}{\omega^{2}\left(  \sigma-1\right)  }\cdot\frac{6\left(
\sigma-1\right)  ^{2}+\omega^{2}}{4\left(  \sigma-1\right)  ^{2}+\omega^{2}}%
\end{equation}
and%
\begin{equation}
S_{01}^{\pi/2}=\frac{8-2\omega^{2}}{T\left(  4+\omega^{2}\right)  ^{2}}%
-\frac{4}{4+\omega^{2}}+\frac{8dT\left[  2\left(  \sigma^{2}+1\right)
+\omega^{2}\right]  }{\left(  \sigma-1\right)  \left(  4+\omega^{2}\right)
\left(  4\sigma^{2}+\omega^{2}\right)  },
\end{equation}
where $\operatorname{sinc}x=\sin\left(  x\right)  /x$. In the following we
will fix $\sigma$ to $\sqrt{2}$ (pump power twice above threshold), as the
results are almost independent of its value as far as it is far enough from
threshold. Hence, the free parameters will be the detection parameters $T$,
$\omega$ and $\varphi$, and the diffusion $d$ which depends on the system parameters.%

\begin{figure}[hb]

\includegraphics[
width=3.4in
]%
{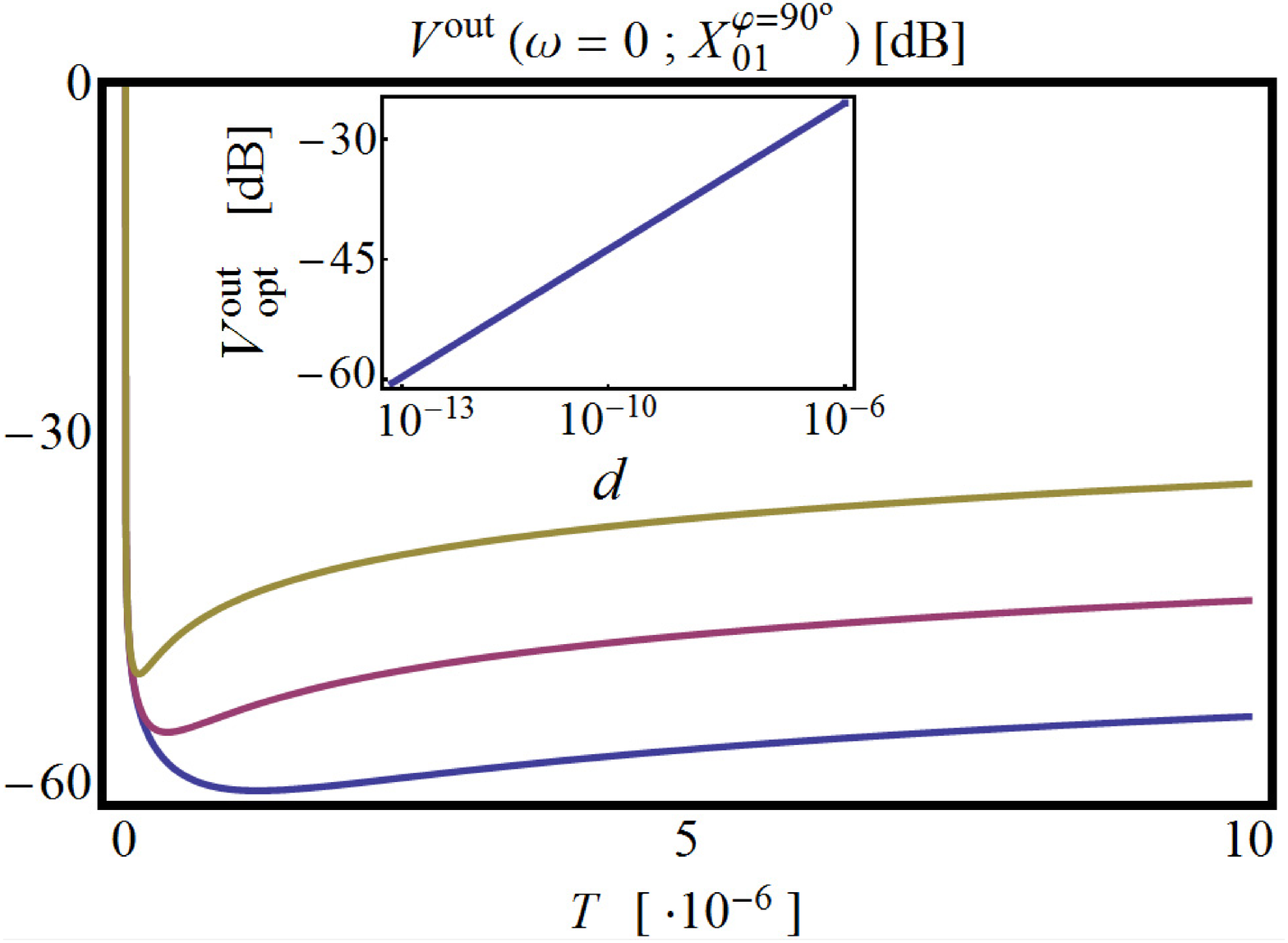}%
\\
{\small Figure 2.- Zero-frequency noise spectrum of the phase quadrature corresponding to a fixed TEM$_{01}$ mode as a function of the detection time $T$. Three different
values of $d$ are considered ($10^{-11}$, $10^{-12}$ and $10^{-13}$ from top to
bottom). The inset shows also this spectrum at zero noise frequency, but evaluated
at the optimum detection time $T_{\mathrm{opt}}$ and as a function of $d$ (note that
the $d$ axis is in logarithmic scale). As mentioned in the text $%
\sigma=\sqrt{2}$.}%

\end{figure}

In this section, the results for $V^{\mathrm{out}}$ are presented in
\textrm{dB} units, defined through the relation $V^{\mathrm{out}}\left[
\mathrm{dB}\right]  =10\log V^{\mathrm{out}}$ (hence, e.g., $-10$ \textrm{dB}
and $-\infty$ \textrm{dB }correspond to 90\% of noise reduction
($V^{\mathrm{out}}=0.1$) and complete noise reduction ($V^{\mathrm{out}}=0$) respectively.)

From expression (\ref{FixedNoiseSpectrum}) we see that the maximum level of
squeezing is obtained at $\omega=0$ (see also Fig. 3a) and when the phase of
the local oscillator is tuned exactly to $\pi/2$. In Fig. 2 we show the noise
spectrum (\ref{FixedNoiseSpectrum}) for these parameters as a function of the
detection time $T$ for 3 different values of $d$. We see that in all cases
there exist an optimum detection time for which squeezing is maximum.
Minimizing Eq. (\ref{FixedNoiseSpectrum}) with $\omega=0$ and $\varphi=\pi/2$
with respect to $T$, it is straightforward to find that this optimum detection
time is given by%
\begin{equation}
T_{\mathrm{opt}}=\sqrt{\frac{\sigma^{2}\left(  \sigma-1\right)  }{d\left(
\sigma^{2}+1\right)  }}, \label{Topt}%
\end{equation}
with an associated noise spectrum $V_{\mathrm{opt}}^{\mathrm{out}%
}=1/T_{\mathrm{opt}}$ (shown in the inset of Fig. 2 as function of $d$).

These results show that large levels of noise reduction are obtained for the
phase quadrature of the fixed TEM$_{01}$ mode, even for values of the
diffusion parameter $d$ as large as $10^{-6}$ (remember that $10^{-13}$ is a
more realistic value). However, in real experiments it is not possible to
ensure that $\varphi=90%
\operatorname{{{}^\circ}}%
$ with an uncertainty below approximately $1.5%
\operatorname{{{}^\circ}}%
$ \cite{Vahlbruch08,Takeno07}, and hence we proceed now to investigate the
level of noise reduction predicted by (\ref{FixedNoiseSpectrum}) when the
local oscillator phase is different from $\varphi=90%
\operatorname{{{}^\circ}}%
$.%

\begin{figure}[hb]

\includegraphics[
width=3.4in
]%
{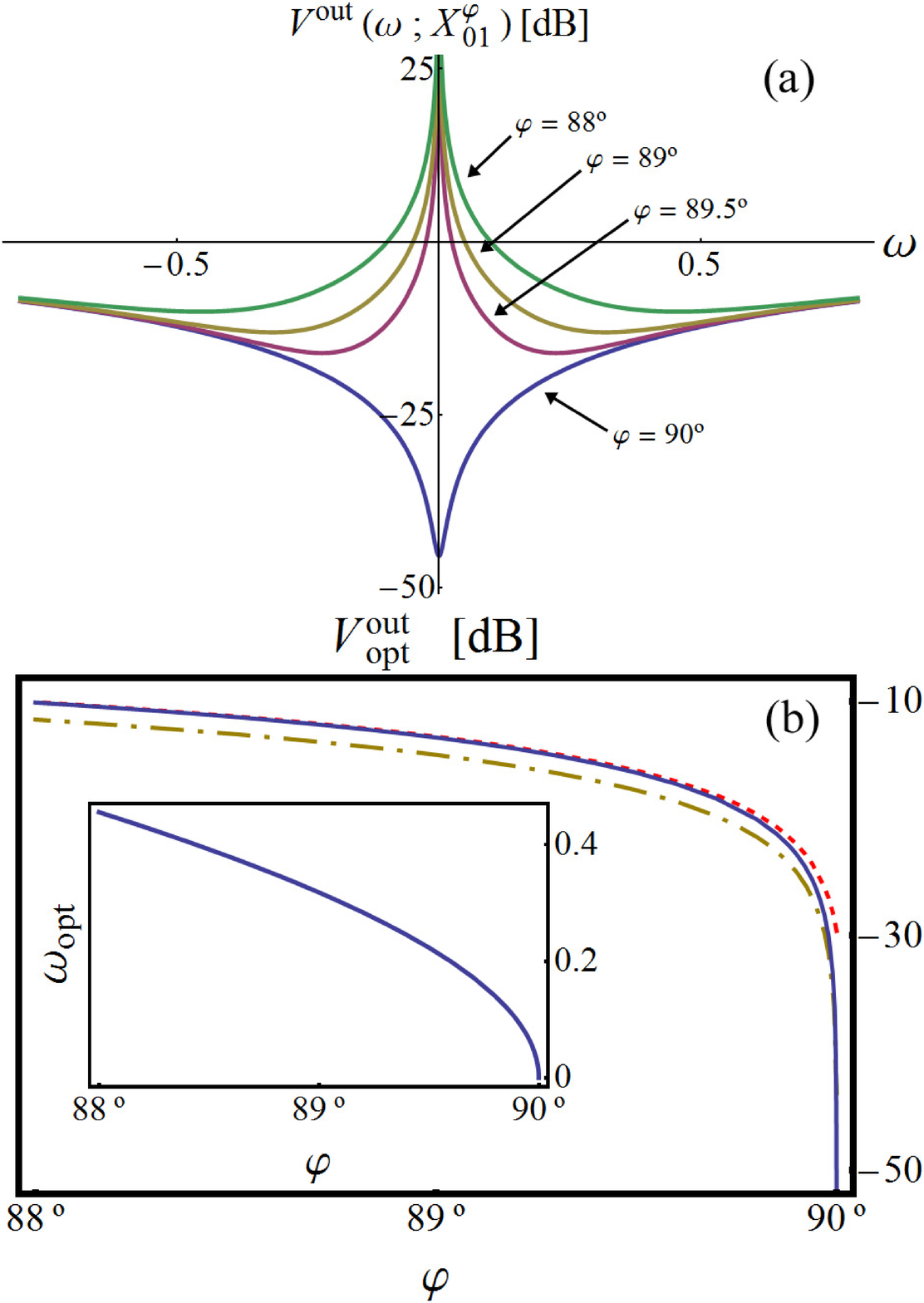}%
\\
{\small Figure 3.- (a) Noise spectrum of the fixed TEM$_{01}$ mode as a function of the
noise frequency $\omega$, and for four different values of the local
oscillator phase ($\varphi=88%
\operatorname{{{}^\circ}}%
$, $89%
\operatorname{{{}^\circ}}%
$, $89.5%
\operatorname{{{}^\circ}}%
$, and $90%
\operatorname{{{}^\circ}}%
$ from top to bottom
curves). It can be appreciated how the infinite fluctuations of $S_{01}^0$ at
zero noise frequency enter in the spectrum for any $\varphi\neq90$. The rest
of parameters are $\sigma=\sqrt{2}$, $d=10^{-10}$, and $T=T_{\mathrm{opt}}$, but the same
behavior appears for any election of the parameters. (b) Noise spectrum of
the fixed TEM$_{01}$ mode evaluated for the optimum parameters $\omega_{\mathrm{opt}}$
and $T_{\mathrm{opt}}$ as a function of $\varphi$ ($d=10^{-13}$ for the blue-solid
curve and $10^{-6}$ for the red-dashed one, having both $\sigma=\sqrt{2}$). In
addition, it is plotted the analogous curve for the single-mode DOPO (gold,
dashed-dotted curve). The inset shows the dependence of the optimum frequency
$\omega_{\mathrm{opt}}$ with the phase of the local oscillator $\varphi$.}%

\end{figure}

Of course, when $\varphi\neq90%
\operatorname{{{}^\circ}}%
$ the noise frequency with maximum squeezing is no longer $\omega=0$, as in
this case the infinite fluctuations of $S_{01}^{0}$ at zero noise frequency,
due to the rotation noise, enter the noise spectrum (see Fig. 3a). By
minimizing $V^{\mathrm{out}}$ with respect to $\omega$ and $T$ for different
values of $\varphi$ and $d$, it is possible to show that the optimum value of
the detection time is almost independent of $\varphi$ for small deviations of
this from $90%
\operatorname{{{}^\circ}}%
$, and hence it is still given to a good approximation by (\ref{Topt}), though
in this case this minimum is less pronounced than in the $\varphi=90%
\operatorname{{{}^\circ}}%
$ case shown in Fig. 2 (i.e., the curve is almost horizontal around
$T_{\mathrm{opt}}$). On the other hand, the optimum noise frequency
$\omega_{\mathrm{opt}}$ is independent of $d$ and depends on $\varphi$ as
shown in the inset of Fig. 3b.

As for the squeezing level, in Fig. 3b we show the noise spectrum evaluated at
$T_{\mathrm{opt}}$ and $\omega_{\mathrm{opt}}$ as a function of $\varphi$ for
2 different values of the diffusion $d$. Together with these curves, we have
plotted the noise spectrum of the single-mode DOPO \cite{DOPOspectrum}
evaluated for its optimum parameters (in this case it is optimized respect to
$\sigma$ and $\omega$) as a function of $\varphi$. We see that the noise
reduction is independent of $d$ as $\varphi$ is taken apart from $90%
\operatorname{{{}^\circ}}%
$. On the other hand, the squeezing level is similar to that of the
single-mode DOPO, as the maximum difference between them are 1.5 \textrm{dB}
(a factor 1.4 in the noise spectrum) in favor of the single-mode DOPO, with
the advantage that in the 2-transverse-mode DOPO this level is independent of
the distance from threshold.

Therefore we see that the phenomenon of non-critical squeezing through
spontaneous rotational symmetry breaking could be observed in the
2-transverse-mode DOPO without the need of following the random rotation of
the generated pattern, which makes its experimental realization feasible with
current available technology.

\section{Beyond the considered approximations}

\subsection{Beyond the adiabatic elimination of the pump}

The first assumption made in the search for the quantum properties of the
system was that $\gamma_{\mathrm{p}}\gg\gamma_{\mathrm{s}}$, a limit that
allowed the adiabatic elimination of the pump field. Now we are going to show
analytically that the phenomenon of squeezing induced by spontaneous
rotational symmetry breaking is still present even without this assumption,
but still working in the domain where linearization is correct.

The way to show this is quite simple; starting from the complete equations
(\ref{ReescaledLangevin}), we expand the amplitudes $\beta_{m}$ around the
classical stationary solution (\ref{AboveThreshold}) as we made in
(\ref{LinExp}), but adding now a similar expression for the pump amplitudes:
$\beta_{0}=1+b_{0}$ and $\beta_{0}^{+}=1+b_{0}^{+}$ (note that for the pump
modes the $b$'s are directly small as the phase of this mode is locked to that
of the injection $\mathcal{E}_{\mathrm{p}}$). Then, linearizing these
equations for the fluctuations and noises, we arrive to a linear system
formally equal to (\ref{LinLan}), but with%
\begin{equation}
\mathbf{b}=%
\begin{pmatrix}
b_{0}\\
b_{0}^{+}\\
b_{+1}\\
b_{+1}^{+}\\
b_{-1}\\
b_{-1}^{+}%
\end{pmatrix}
\text{, }\boldsymbol{\zeta}=%
\begin{pmatrix}
0\\
0\\
\zeta\left(  \tau\right) \\
\zeta^{+}\left(  \tau\right) \\
\zeta^{\ast}\left(  \tau\right) \\
\left[  \zeta^{+}\left(  \tau\right)  \right]  ^{\ast}%
\end{pmatrix}
,
\end{equation}
and a linear matrix%
\begin{equation}
\mathcal{L}=%
\begin{pmatrix}
-\kappa & 0 & -\rho & 0 & -\rho & 0\\
0 & -\kappa & 0 & -\rho & 0 & -\rho\\
\rho & 0 & -1 & 0 & 0 & 1\\
0 & \rho & 0 & -1 & 1 & 0\\
\rho & 0 & 0 & 1 & -1 & 0\\
0 & \rho & 1 & 0 & 0 & -1
\end{pmatrix}
.
\end{equation}
Although in this case $\mathcal{L}$ is not hermitian, it can be checked that
it possess a biorthonormal basis \cite{Biorthonormal}, and in particular, the
following two vectors are present in its eigensystem: $\mathbf{w}_{0}^{\prime
}=\frac{1}{2}\operatorname{col}\left(  0,0,1,-1,-1,1\right)  $ and
$\mathbf{w}_{1}^{\prime}=\frac{1}{2}\operatorname{col}\left(
0,0,1,1,-1,-1\right)  $, with corresponding eigenvalues $\lambda_{0}^{\prime
}=0$ and $\lambda_{1}^{\prime}=-2$. These eigenvectors have null projection
onto the pump subspace, and coincide with $\mathbf{w}_{0}$ and $\mathbf{w}%
_{1}$ in what concerns to the signal subspace (\ref{Eigensystem}). Hence, all
the properties derived from these vectors are still present without any
change. In particular, as they account for the diffusion of $\theta$ and the
squeezing properties of the dark mode, we can conclude that these properties
are still present when working out of the limit $\gamma_{\mathrm{p}}\gg
\gamma_{\mathrm{s}}$.

\subsection{Numerical simulation of the nonlinear equations}

In this section we will show that the diffusion of the orientation and the
associated non-critical squeezing of the dark mode, which have been found by
linearizing the Langevin equations, are also present when we consider the full
nonlinear problem. To do so, we will solve numerically the complete stochastic
equations (\ref{ReescaledLangevin}) using the semi-implicit algorithm
developed by Drummond and Mortimer in \cite{Drummond91}.

The details of the numerical simulation are explained in Appendix F. Here we
just want to point out that the important parameters of the simulation are the
step size $\Delta\tau$ used to arrive from $\tau=0$ to the final integration
time $\tau_{\mathrm{end}}$, and the number of stochastic trajectories, say
$\Sigma$, which are used to evaluate stochastic averages. The initial
conditions $\beta_{m}\left(  0\right)  $ are not important as the results in
the stationary limit are independent of them. The system parameters which have
been chosen for the simulation are $\sigma=\sqrt{2}$, $\kappa=1$ (to show also
numerically that the adiabatic elimination has nothing to do with the
phenomenon), and $g=10^{-3}$. We haven't chosen a smaller value for $g$ (like
$10^{-6}$ as followed from the physical parameters considered in Appendix A)
because such a small number can make the simulation fail; nevertheless all the
results we are going to show should be independent of $g$ and $\sigma$, and we have
also tested that the same results are obtained for other values of these.%

\begin{figure}[ht]

\includegraphics[
width=3.3in
]%
{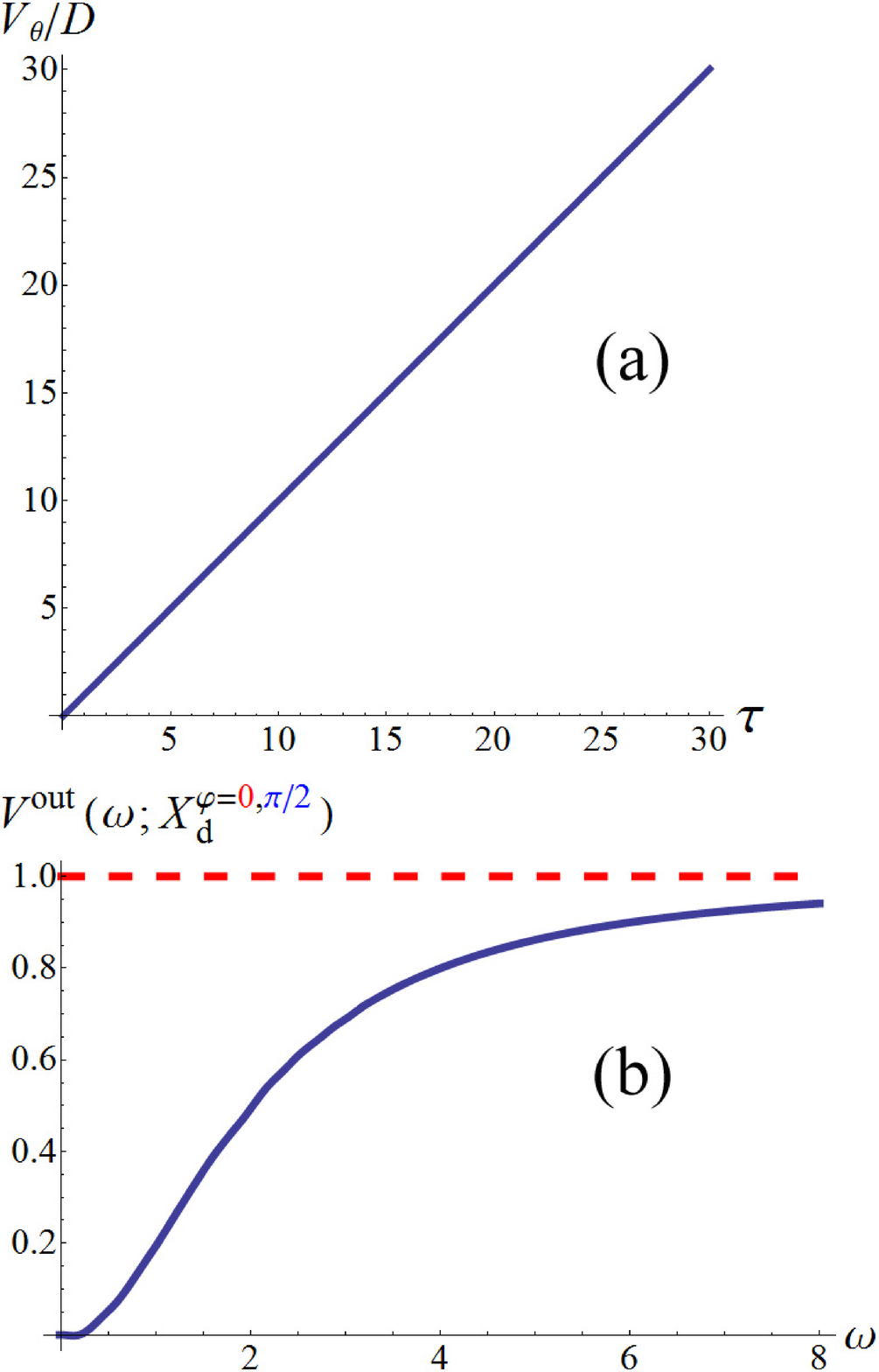}%
\\
{\small Figure 4.- (a) Evolution of the variance of the orientation $\theta$ given
by the numerical simulation. It has been divided by the slope $D$ predicted
by the linearized theory (\ref{Theta Var}), so the straight line obtained is
in perfect agreement with this linear result. Quantitatively, a linear
regression analysis shows that the slope obtained by the data is $1.00108$,
with standard error $9\cdot10^{-5}$. (b) Noise spectrum of the amplitude
(red-dashed curve) and phase (blue-solid curve) quadratures of the dark mode
as obtained by the numerical simulation. The results are in perfect agreement
with the ones predicted by the linearized theory (\ref{NoiseSpectra}). A
nonlinear regression analysis of the numerical data respect to the function
$a\left(  \omega/2\right)  ^2/\left[  b+a\left(  \omega/2\right)  ^2\right]
$, shows that $\left(  a,b\right)  =\left(  0.9446,0.9591\right)  $ are the
best fit parameters having both standard errors $4\cdot10^{-4}$, which is in
good agreement with the values $\left(  a,b\right)  =\left(  1,1\right)  $
predicted by (\ref{NoiseSpectra}).}%

\end{figure}

It is also important to note that we have defined a general quadrature of the
rotating dark mode as (directly from (\ref{LGtoHGs}) with $\beta=\theta$)%
\begin{align}
X_{\mathrm{d}}^{\varphi}  &  =\frac{i}{\sqrt{2}}\left[  e^{-i\varphi}\left(
e^{i\theta}\beta_{+1}-e^{-i\theta}\beta_{-1}\right)  \right] \\
&  -\frac{i}{\sqrt{2}}\left[  e^{i\varphi}\left(  e^{-i\theta}\beta_{+1}%
^{+}-e^{i\theta}\beta_{-1}^{+}\right)  \right]  ,\nonumber
\end{align}
with $\theta$ defined within the positive \textit{P} representation through%
\begin{equation}
e^{2i\theta}=\frac{\beta_{-1}\beta_{+1}^{+}}{\left\vert \beta_{-1}\right\vert
\left\vert \beta_{+1}^{+}\right\vert }.
\end{equation}

Now let us show the results evaluated for the following simulation parameters:
$\Delta\tau=3\cdot10^{-3}$, $\tau_{\mathrm{end}}=30$ and $\Sigma=10^{7}$. This
simulation has been compared with other ones having different values of these
parameters to ensure convergence.

In Fig. 4a we show the variance of $\theta$ as a function of time. The result
has been normalized to $D$, so that the linear result (\ref{Theta Var})
predicts a straight line forming 45$%
\operatorname{{{}^\circ}}%
$ with respect to the time axis. It can be appreciated that this is indeed
what shows the simulation.

In Fig. 4b, we show the numerical results for the noise spectrum associated to
the quadratures of the dark mode. Only times above $\tau=10$ have been
considered in the correlation function to ensure being working in the
stationary limit. Again, the results shown in Fig. 4b are in perfect agreement
with the linear predictions (\ref{XsSpectrum}) and (\ref{YsSpectrum}).

These results show that the phenomenon of non-critically squeezed light via
spontaneous rotational symmetry breaking is not a product of the linearization.

\section{Conclusions}

In conclusion, we have developed in detail the quantum theory of the
2-transverse-mode DOPO introduced in \cite{PRL}. We have studied some
important features not considered in that reference:

\begin{itemize}
\item When pumping with a gaussian mode a rotationally symmetric DOPO properly
tuned to the first family of transverse modes at the signal frequency,
classical emission takes place in a TEM$_{10}$ mode with an arbitrary
orientation in the transverse plane. Hence, once the threshold is crossed, the
rotational symmetry is spontaneously broken, and we can talk about a bright
mode (the generated one which breaks the symmetry) and a dark mode (the mode
orthogonal to the generated one).

\item The symmetry breaking reflects in the appearance of a Goldstone mode in
the matrix governing the linear evolution of the fluctuations above threshold.
The null eigenvalue of this mode allows quantum noise to change the
orientation of the bright mode randomly. Though continuously increasing with
time, this rotation of the classically excited pattern is quite slow when
working above threshold.

\item As for the squeezing properties, it has been proved that the bright mode
has the same behavior as the single-mode DOPO, i.e., perfect squeezing
appears only at threshold (within the linearized theory) and degrades fast as
pump is moved apart from this level. On the other hand, accompanying the
Goldstone mode it appears another mode whose associated eigenvalue takes the
minimum possible value $\left(  -2\right)  $. These modes are responsible of
the remarkable properties of the dark mode: Its phase quadrature is perfectly
squeezed at any pump level, while its amplitude quadrature carries only with
vacuum fluctuations \cite{PRL} (in apparent violation of the uncertainty
principle). A simple explanation of this phenomenon in terms of "angle -
angular momentum" uncertainty relation appears once one notices that the dark
mode coincides (up to a $\pi/2$ phase) with the OAM of the generated pattern.

\item We have proved that the apparent violation of the uncertainty principle
is just that, apparent, as the conjugate pair of the squeezed quadrature is
not another quadrature but the orientation of the bright mode, which in fact
is completely undetermined in the long term.

\item Next we have pointed out that in order to measure the quantum properties
of the dark mode, one has to use a TEM$_{10}$ local oscillator that is
perfectly matched to the orientation of this mode at any time. However, the
mode is rotating randomly, which seems to make impossible the perfect
matching. For this reason we have studied the situation in which the local
oscillator is matched to the dark mode's orientation only at the initial time,
remaining fixed during the detection time. We have shown that arbitrarily
large levels of noise reduction can be obtained even in this case if the phase
of the local oscillator is exactly $\pi/2$. We then considered phase
deviations up to 2$%
\operatorname{{{}^\circ}}%
$ (1.5$%
\operatorname{{{}^\circ}}%
$ seems to be the current experimental limit \cite{Vahlbruch08,Takeno07}),
comparing the results with that predicted for the single-mode DOPO; similar
levels are obtained for both, with the advantage that in the
2-transverse-mode DOPO this level is independent of the distance from
threshold, and hence, non-critical.

\item In the last part of the article we have shown that the assumptions made
in order to analytically solve the problem are not the responsible for the
quantum properties of the dark mode. In particular, we have shown that when the pump is not adiabatically eliminated, the
Goldstone mode and its companion with the lowest possible eigenvalue remain unchanged
in the matrix governing the linear evolution of the full problem, and hence all the properties derived from them are still present. Finally, we have used numerical simulations to show that the predictions of
the linearized equations are in perfect agreement with those of the full
nonlinear equations, and hence the perfect, non-critical squeezing of the dark
mode is not a by-product of the linearization.
\end{itemize}

We believe that the analyses presented in this paper (together with the fact
that the properties of the dark mode are not too sensitive to imperfections in
the rotational symmetry of the DOPO \cite{PRL}, and that the phenomenon is
present in other kinds of nonlinear resonators \cite{OtherUs}) show that that
the phenomenon of non-critical squeezing induced by spontaneous rotational
symmetry breaking is a robust phenomenon.

We thank Ferran V. Garcia--Ferrer for his help in Section V.A. This work has
been supported by the Spanish Ministerio de educaci\'{o}n y Ciencia and the
European Union FEDER through Project FIS2008-06024-C03-01. C N-B is
a grant holder of the FPU programme of the Ministerio de Educaci\'{o}n y
Ciencia (Spain). A.R. acknowledges financial support from the Universitat de Val\`{e}ncia through its program
``Convocatoria de Estancias Temporales para Investigadores Invitados".

\appendix

\section{Model and physical parameters}

During the article some parameters have been used to model the DOPO. Here we
want to give explicit expressions for them in terms of physical quantities for
the case of a DOPO having a Fabry-Perot cavity of effective length $L$ and
formed by two identical spherical mirrors with curvature radii $R$ for
simplicity. The nonlinear crystal is placed at the waist plane of this cavity
and has refractive index $n$, second order susceptibility $\chi^{\left(
2\right)  }$ and axial length $l$ (assumed to be much smaller than the
Rayleigh length of the cavity).

For this cavity configuration, the beam radius at the waist plane of the
resonator is given by \cite{Hodgson05}%
\begin{equation}
w_{j}^{2}=\frac{\lambda_{j}L}{2\pi}\sqrt{\frac{2R}{L}-1},\text{ }%
j=\mathrm{p},\mathrm{s},
\end{equation}
where $\lambda_{j}$ is the wavelength of the considered mode inside the cavity.

On the other hand, the parameters which appear in the Hamiltonian
(\ref{Hamiltonian}) accounting for the external laser pump and the nonlinear
interaction inside the crystal, namely $\mathcal{E}_{\mathrm{p}}$ and $\chi$,
respectively have the following expression:%
\begin{align}
\mathcal{E}_{\mathrm{p}}  &  =\sqrt{\frac{n\lambda_{\mathrm{p}}\gamma
_{\mathrm{p}}}{2\pi\hbar c}P_{\mathrm{laser}}}\\
\chi &  =\frac{3\pi\chi^{\left(  2\right)  }l}{w_{\mathrm{p}}}\sqrt
{\frac{\hbar}{\varepsilon_{0}}}\left(  \frac{c}{nL\lambda_{\mathrm{p}}%
}\right)  ^{3/2},\nonumber
\end{align}
being $\gamma_{j}=c\mathcal{T}_{j}/2L$ the cavity decay rate at the considered
frequency ($\mathcal{T}_{j}$ is the corresponding transmission factor through
the input mirror), and $P_{\mathrm{laser}}$ the power of the injected laser.
In order to obtain these expressions all the transmitted pump power is assumed
to be focalized inside the transverse dimensions of the nonlinear crystal.

Through the article, some expressions have been evaluated for concrete system
parameters. We have taken as typical parameters the following ones%
\[%
\begin{array}
[c]{llll}%
\lambda_{\mathrm{p}}=400\text{ }%
\operatorname{nm}%
, & R=1\text{ }%
\operatorname{m}%
, & \mathcal{T}_{\mathrm{p}}=0.1, & \chi^{\left(  2\right)  }=2\text{ }\frac{%
\operatorname{pm}%
}{%
\operatorname{V}%
},\\
L=0.1\text{ }%
\operatorname{m}%
, & l=1\text{ }%
\operatorname{mm}%
, & \mathcal{T}_{\mathrm{s}}=0.01, & n=2.5,
\end{array}
\]
leading to the following model parameters%
\[%
\begin{array}
[c]{ll}%
w_{\mathrm{p}}=167\text{ }%
\operatorname{\mu m}%
, & \chi=64\text{ }%
\operatorname{s}%
^{-1},\\
\gamma_{\mathrm{p}}=0.15\text{ }%
\operatorname{ns}%
^{-1}, & \gamma_{\mathrm{s}}=0.015\text{ }%
\operatorname{ns}%
^{-1}.
\end{array}
\]

\section{Connection between Fokker-Planck and Langevin
equations}

During the article we make extensive use of the equivalence between Langevin
and Fokker-Planck equations. In this appendix we want to briefly review this connection.

Consider a set of real variables $\mathbf{x}=\left(  x_{1},x_{2}%
,...,x_{n}\right)  $, satisfying stochastic Langevin equations%
\begin{equation}
\frac{d\mathbf{x}}{dt}=\mathbf{A}\left(  \mathbf{x}\right)  +\mathcal{B}%
\left(  \mathbf{x}\right)  \boldsymbol{\eta}\left(  t\right)  \text{,}
\label{GenLan}%
\end{equation}
where $\mathbf{A}\left(  \mathbf{x}\right)  $ and $\mathcal{B}\left(
\mathbf{x}\right)  $ are a vector and a matrix which depend on the variables,
and the components of $\boldsymbol{\eta}\left(  t\right)  $ are real noises
satisfying the usual statistical properties (\ref{RealNoiseStat}).

The theory of stochastic processes \cite{Gardiner09} states that the
stochastic average of any function of the variables $f\left(  \mathbf{x}%
\right)  $ can be evaluated as%
\begin{equation}
\left\langle f\left(  \mathbf{x}\right)  \right\rangle _{\mathrm{stochastic}%
}=\int d^{n}\mathbf{x}P\left(  \mathbf{x}\right)  f\left(  \mathbf{x}\right)
,
\end{equation}
where the probability distribution $P\left(  \mathbf{x}\right)  $ satisfies
the Fokker-Planck equation%
\begin{equation}
\partial_{\tau}P\left(  \tau;\mathbf{x}\right)  =\left[  -\sum_{i}\partial
_{i}A_{i}^{\left(  FP\right)  }+\frac{1}{2}\sum_{i,j}\partial_{i,j}%
^{2}\mathcal{D}_{ij}\right]  P\left(  \tau;\mathbf{x}\right)  \text{,}%
\end{equation}
having drift vector and diffusion matrix%
\begin{equation}
A_{i}^{\left(  FP\right)  }=A_{i}+\frac{\nu}{2}\sum_{jk}\mathcal{B}%
_{kj}\partial_{k}\mathcal{B}_{ij},
\end{equation}
and%
\begin{equation}
\mathcal{D}=\mathcal{BB}^{T},
\end{equation}
respectively.

The parameter $\nu$ is $0$ or $1$ depending on whether we interpret Eqs.
(\ref{GenLan}) as Ito or Stratonovich stochastic equations. In the current
article any stochastic equation is interpreted \`{a} la Stratonovich, which
allows us using the usual rules of calculus. Note that if the extra term
involving derivatives of the noise matrix is zero, Ito and Stratonovich
interpretations are equivalent, what happens for example on Eqs.
(\ref{Langevin}) but not on Eqs. (\ref{AdElEqs}).

\section{Considerations on the linearization procedure}

Following the linearization procedure we explained in Section III.A leads not
to Eq. (\ref{LinLan}) directly, but to the following one%
\begin{equation}
i\left(  \mathcal{G}\mathbf{b}-2\rho\mathbf{w}_{0}\right)  \dot{\theta
}+\mathbf{\dot{b}=}\mathcal{L}\mathbf{b}+g\mathcal{K}\left(  \theta\right)
\boldsymbol{\zeta}\left(  \tau\right)  \text{,}\label{FullLinLan}%
\end{equation}
where%
\begin{align*}
\mathcal{K}\left(  \theta\right)   &  =\operatorname{diag}\left(  e^{i\theta
},e^{-i\theta},e^{-i\theta},e^{i\theta}\right)  \text{,}\\
\mathcal{G} &  =\operatorname{diag}\left(  -1,1,1,-1\right)  ,
\end{align*}
and the rest of vectors and symbols were defined in the corresponding section.
Note that the differences between this system of equations and the one used in
the text (\ref{LinLan}) are the matrix $\mathcal{K}\left(  \theta\right)  $
and the $i\mathcal{G}\mathbf{b}\dot{\theta}$ term. The latter is of order
$g^{2}$ (as the $b$'s and $\dot{\theta}$ are of order $g$), and hence it can
be simply removed within the linearized theory.

Understanding why $\mathcal{K}\left(  \theta\right)  $ can be removed from the
linearized equations is a little more involved. Projecting these equations
onto the eigensystem of $\mathcal{L}$ (\ref{Eigensystem}) and defining the
vector $\mathbf{c}=\operatorname{col}\left(  \theta,c_{1},c_{2},c_{3}\right)
$ leads to the following system of equations (remember that we set $c_{0}=0$)
\begin{equation}
\mathbf{\dot{c}=-}\Lambda\mathbf{c}+g\mathcal{BR}\left(  \theta\right)
\boldsymbol{\eta}\left(  \tau\right)  \label{cLinLan}%
\end{equation}
with%
\begin{align*}
\Lambda &  =2\operatorname{diag}\left(  0,1,\sigma-1,\sigma\right)  ,\\
\mathcal{B} &  =\operatorname{diag}\left(  1/2\rho,i,1,1\right)  ,\\
\mathcal{R}\left(  \theta\right)   &  =\mathcal{R}_{1,3}\left(  \theta\right)
\mathcal{R}_{2,4}\left(  -\theta\right)  ,
\end{align*}
being $\mathcal{R}_{i,j}\left(  \theta\right)  $ the 2-dimensional rotation
matrix of angle $\theta$ acting on the $i-j$ subspace, and where the
components of vector $\boldsymbol{\eta}\left(  \tau\right)  $ are real,
independent noises satisfying the usual statistical properties
(\ref{RealNoiseStat}). Now, we will prove that this system, and the same with
$\mathcal{R}\left(  \theta=0\right)  $ are equivalent within the linearized
theory, and hence (\ref{FullLinLan}) and (\ref{LinLan}) are equivalent too.

To show this, we just write the Fokker-Planck equation corresponding to this
stochastic system (see Appendix B), whose drift vector and diffusion matrix
are found to be%
\begin{equation}
\mathbf{A}=\Lambda\mathbf{c}+\frac{g^{2}}{4\rho}\operatorname{col}\left(
0,0,1,0\right)  ,
\end{equation}
and%
\begin{equation}
\mathcal{D}=g^{2}\mathcal{BB}^{T},
\end{equation}
respectively. Note that in the last equation we have used that $\mathcal{R}%
\left(  \theta\right)  $ is an orthogonal matrix.

The proof is completed by writing the stochastic system corresponding to this
Fokker-Planck equation up to the linear order in $g$, which reads%
\begin{equation}
\mathbf{\dot{c}=-}\Lambda\mathbf{c}+g\mathcal{B}\boldsymbol{\eta}\left(
\tau\right)  ,
\end{equation}
corresponding to (\ref{cLinLan}) with $\mathcal{R}\left(  \theta=0\right)  $
as we wanted to prove.

Hence, removing $\mathcal{K}\left(  \theta\right)  $ and neglecting the
$i\mathcal{G}\mathbf{b}\dot{\theta}$ term\ from (\ref{FullLinLan}) doesn't
change its equivalent Fokker-Planck equation within the linearized
description, and thus equation (\ref{LinLan}) must lead to the same
predictions as (\ref{FullLinLan}).

\section{Linear Langevin equations: solution, correlation and
spectrum}

In this appendix we solve the linear evolution equations (\ref{ProjLinLan}) of
the projections $c_{m}$, finding their 2-time-correlation function and the
associated spectrum. These equations are of the general type%
\begin{equation}
\dot{c}=-\lambda c+\Gamma\eta\left(  \tau\right)  ,
\end{equation}
where $\eta\left(  \tau\right)  $ is a real noise satisfying the usual white
noise statistics (\ref{RealNoiseStat}), $\lambda$ is a real, positive
parameter, and $\Gamma$ is a parameter that might be complex.

By making the variable change $z\left(  \tau\right)  =c\left(  \tau\right)
\exp\left(  \lambda\tau\right)  $, and considering times larger that
$\lambda^{-1}$ (stationary limit), it is straightforward to find the following
solution%
\begin{equation}
c\left(  \tau\right)  =\Gamma\int_{0}^{\tau}d\tau_{1}\eta\left(  \tau
_{1}\right)  e^{\lambda\left(  \tau_{1}-\tau\right)  }.
\end{equation}

From the statistical properties of noise (\ref{RealNoiseStat}) we see that
this solution has zero mean, $\left\langle c\left(  \tau\right)  \right\rangle
=0$, and correlation%
\begin{align}
\left\langle c\left(  \tau\right)  c\left(  \tau^{\prime}\right)
\right\rangle  &  =\Gamma^{2}e^{-\lambda\left(  \tau+\tau^{\prime}\right)  }\\
&  \times\int_{0}^{\tau}d\tau_{1}\int_{0}^{\tau^{\prime}}d\tau_{2}%
\delta\left(  \tau_{1}-\tau_{2}\right)  e^{\lambda\left(  \tau_{1}+\tau
_{2}\right)  }.\nonumber
\end{align}
Considering separately the cases $\tau^{\prime}>\tau$ and $\tau^{\prime}<\tau
$, this integral is easily carried out, yielding (again the limit $\tau
\gg\lambda^{-1}$ is considered)%
\begin{equation}
\left\langle c\left(  \tau\right)  c\left(  \tau^{\prime}\right)
\right\rangle =\frac{\Gamma^{2}}{2\lambda}e^{-\lambda\left\vert \tau^{\prime
}-\tau\right\vert },
\end{equation}
where we see that this function depends only on the time difference
$\left\vert \tau^{\prime}-\tau\right\vert $, what justifies the name
\textquotedblleft stationary limit\textquotedblright\ for the $\tau\gg
\lambda^{-1}$ approximation.

Finally, the spectrum of this correlation is found to be%
\begin{equation}
\tilde{C}\left(  \omega\right)  =\int_{-\infty}^{+\infty}d\bar{\tau
}e^{-i\omega\bar{\tau}}\left\langle c\left(  \tau\right)  c\left(
\tau+\bar{\tau}\right)  \right\rangle =\frac{\Gamma^{2}}{\lambda^{2}%
+\omega^{2}}.
\end{equation}

Hence, particularized to the equations for the projections $c_{m}$
(\ref{ProjLinLan}), the correlations read%
\begin{align}
\left\langle c_{1}\left(  \tau\right)  c_{1}\left(  \tau^{\prime}\right)
\right\rangle  &  =-\frac{g^{2}}{4}e^{-2\left\vert \tau^{\prime}%
-\tau\right\vert }\label{Ccorr}\\
\left\langle c_{2}\left(  \tau\right)  c_{2}\left(  \tau^{\prime}\right)
\right\rangle  &  =\frac{g^{2}}{4\left(  \sigma-1\right)  }e^{-2\left(
\sigma-1\right)  \left\vert \tau^{\prime}-\tau\right\vert }\nonumber\\
\left\langle c_{3}\left(  \tau\right)  c_{3}\left(  \tau^{\prime}\right)
\right\rangle  &  =\frac{g^{2}}{4\sigma}e^{-2\sigma\left\vert \tau^{\prime
}-\tau\right\vert },\nonumber
\end{align}
while the spectra read%
\begin{align}
\tilde{C}_{1}\left(  \omega\right)   &  =-\frac{g^{2}}{4+\omega^{2}%
}\label{CSpectrum}\\
\tilde{C}_{2}\left(  \omega\right)   &  =\frac{g^{2}}{4\left(  \sigma
-1\right)  ^{2}+\omega^{2}}\nonumber\\
\tilde{C}_{3}\left(  \omega\right)   &  =\frac{g^{2}}{4\sigma^{2}+\omega^{2}%
},\nonumber
\end{align}
and the stationary limit is reached when $\tau\gg0.5$, $\tau\gg0.5\left(
\sigma-1\right)  ^{2}$ and $\tau\gg0.5\sigma^{-1}$ respectively.

\section{Calculating correlations for the undamped orientation
$\theta$.}

In this appendix we show how $S\left(  \tau_{1},\tau_{2}\right)  $ and
$C\left(  \tau_{1},\tau_{2}\right)  $ defined in (\ref{StillCorr}) can be
evaluated from the evolution equation of $\theta$ (\ref{ThetaEvo}). As clearly
explained in \cite{MandelWolf}, this equation defines a Wiener or Random Walk
process equivalently described by the following Fokker-Planck equation:%
\begin{equation}
\partial_{\tau}P\left(  \theta,\tau\right)  =\frac{1}{2}D\partial_{\theta}%
^{2}P\left(  \theta,\tau\right)  .
\end{equation}
In the same reference, it is also proved that the 2-time joint probability
associated to this simple diffusion equation is%
\begin{equation}
P\left(  \theta_{2},\tau_{2};\theta_{1},\tau_{1}\right)  =\frac{1}{2\pi
D\sqrt{\tau_{1}\left(  \tau_{2}-\tau_{1}\right)  }}e^{-\frac{1}{2D}\left[
\frac{\left(  \theta_{2}-\theta_{1}\right)  ^{2}}{\tau_{2}-\tau_{1}}%
+\frac{\theta_{1}^{2}}{\tau_{1}}\right]  },
\end{equation}
which gives us the probability of passing from orientation $\theta_{1}$ at time
$\tau_{1}$, to $\theta_{2}$ at time $\tau_{2}$ (it is assumed that $\theta=0$
at the initial time $\tau=0$). Of course, in this expression $\tau_{2}%
>\tau_{1}$. From this joint probability the correlations are calculated as the
integrals%
\begin{align}
S\left(  \tau_{1},\tau_{2}\right)   &  =%
{\displaystyle\iint\limits_{-\infty}^{+\infty}}
d\theta_{1}d\theta_{2}P\left(  \theta_{2},\tau_{2};\theta_{1},\tau_{1}\right)
\sin\theta_{1}\sin\theta_{2}\\
C\left(  \tau_{1},\tau_{2}\right)   &  =%
{\displaystyle\iint\limits_{-\infty}^{+\infty}}
d\theta_{1}d\theta_{2}P\left(  \theta_{2},\tau_{2};\theta_{1},\tau_{1}\right)
\cos\theta_{1}\cos\theta_{2}.\nonumber
\end{align}
These integrals are easily evaluated, yielding%
\begin{equation}
S\left(  \tau_{1},\tau_{2}\right)  =e^{-\frac{1}{2}D\left(  \tau_{1}+\tau
_{2}\right)  }\sinh\left[  D\min\left(  \tau_{1},\tau_{2}\right)  \right]  ,
\label{SinCorr}%
\end{equation}
and%
\begin{equation}
C\left(  \tau_{1},\tau_{2}\right)  =e^{-\frac{1}{2}D\left(  \tau_{1}+\tau
_{2}\right)  }\cosh\left[  D\min\left(  \tau_{1},\tau_{2}\right)  \right]  .
\label{CosCorr}%
\end{equation}

\section{Details of the numerical simulation}

In this last appendix we want to briefly resume the details concerning the
numerical simulation of the Langevin equations (\ref{ReescaledLangevin}).

The first important property of these equations is that, irrespective of the
initial conditions, the amplitudes corresponding to opposite OAM modes become
complex-conjugate after a short transitory time, i.e., $\left(  \beta
_{-1},\beta_{-1}^{+}\right)  \rightarrow\left(  \beta_{+1}^{\ast},\left[
\beta_{+1}^{+}\right]  ^{\ast}\right)  $. Hence, if the initial conditions are
chosen so that the OAM pairs are complex-conjugate, we can be sure that they
will remain complex-conjugate during the evolution. In particular, we have
chosen the above threshold stationary solution (\ref{AboveThreshold}) with
$\theta=0$ as the initial condition. Under these conditions, the 6 Langevin
equations (\ref{ReescaledLangevin}) reduce to the following 4 (which we write in matrix
form):%
\begin{equation}
\boldsymbol{\dot{\beta}}=\mathbf{A}\left(  \boldsymbol{\beta}\right)
+\mathcal{B}\left(  \boldsymbol{\beta}\right)  \cdot\boldsymbol{\zeta}\left(
\tau\right)  \text{,} \label{ReducedEqs}%
\end{equation}
with%
\begin{align}
\boldsymbol{\beta}  &  =%
\begin{pmatrix}
\beta_{0}\\
\beta_{0}^{+}\\
\beta_{+1}\\
\beta_{+1}^{+}%
\end{pmatrix}
,\text{ }\boldsymbol{\zeta}\left(  \tau\right)  =%
\begin{pmatrix}
0\\
0\\
\zeta\left(  \tau\right) \\
\zeta^{+}\left(  \tau\right)
\end{pmatrix}
,\\
\mathbf{A}\left(  \boldsymbol{\beta}\right)   &  =%
\begin{pmatrix}
\sigma-\beta_{0}-\left\vert \beta_{+1}\right\vert ^{2}\\
\sigma-\beta_{0}^{+}-\left\vert \beta_{+1}^{+}\right\vert ^{2}\\
-\beta_{+1}+\beta_{0}\left[  \beta_{+1}^{+}\right]  ^{\ast}\\
-\beta_{+1}^{+}+\beta_{0}^{+}\beta_{+1}^{\ast}%
\end{pmatrix}
,\nonumber\\
\mathcal{B}\left(  \boldsymbol{\beta}\right)   &  =g\operatorname{diag}\left(
0,0,\sqrt{\beta_{0}},\sqrt{\beta_{0}^{+}}\right)  .\nonumber
\end{align}

In order to solve numerically these equations we use the semi-implicit
algorithm developed in Ref. \cite{Drummond91}. This algorithm is a
finite-differences based method in which the total integration time
$\tau_{\mathrm{end}}$ (the integration is supposed to begin always at $\tau
=0$) is divided in $N$ segments, creating hence a lattice of times $\left\{
\tau_{n}\right\}  _{n=0,1,...,N}$ separated by time steps $\Delta\tau
=\tau_{\mathrm{end}}/N$. Then, a recursive algorithm starts in which the
amplitudes at time $\tau_{n}$, say $\boldsymbol{\beta}_{n}$, are found from
the amplitudes $\boldsymbol{\beta}$$^{n-1}$ at an earlier time $\tau_{n-1}$
from%
\begin{equation}
\boldsymbol{\beta}^{n}=\boldsymbol{\beta}^{n-1}+\Delta\tau\mathbf{A}\left(
\boldsymbol{\tilde{\beta}}^{n}\right)  +\mathcal{B}\left(  \boldsymbol{\tilde{\beta}%
}^{n}\right)  \cdot\mathbf{W}^{n},
\end{equation}
where $\boldsymbol{\tilde{\beta}}$$^{n}$ is an approximation to the amplitudes
at the mid-point between $\tau_{n-1}$ and $\tau_{n}$ (hence the name
\textquotedblleft semi-implicit\textquotedblright\ for the algorithm) and the
components of $\mathbf{W}^{n}$ are independent discrete noises $W_{j}^{n}$
with null mean and satisfying the correlations%
\begin{equation}
\left\langle W_{j}^{m},\left[  W_{k}^{n}\right]  ^{\ast}\right\rangle
=\Delta\tau\delta_{mn}\delta_{jk}.
\end{equation}

The mid-point approximation is found from the following iterative algorithm%
\begin{equation}
\boldsymbol{\tilde{\beta}}^{n,p}=\boldsymbol{\beta}^{n-1}+\frac{1}{2}\left[
\Delta\tau\mathbf{A}\left(  \boldsymbol{\tilde{\beta}}^{n,p-1}\right)
+\mathcal{B}\left(  \boldsymbol{\tilde{\beta}}^{n,p-1}\right)  \cdot\mathbf{W}%
^{n}\right]  ,
\end{equation}
where $\boldsymbol{\tilde{\beta}}$$^{n,0}=$$\boldsymbol{\beta}$$^{n-1}$, being
$p$ the iteration index (two iterations are carried in our simulations), while
the discrete noises can be simulated at any step as \cite{Fox88}%
\begin{equation}
W_{j}^{n}=\sqrt{\Delta\tau}\left[  r\left(  z_{j},z_{j}^{\prime}\right)
+ir\left(  y_{j},y_{j}^{\prime}\right)  \right]  ,
\end{equation}
with%
\begin{equation}
r\left(  z,z^{\prime}\right)  =\sqrt{-\log z}\cos\left(  2\pi z^{\prime
}\right)  ,
\end{equation}
being $z_{j}$, $z_{j}^{\prime}$, $y_{j}$ and $y_{j}^{\prime}$ independent
random numbers uniformly distributed along the interval $\left[  0,1\right]  $.

This algorithm allows us to simulate one stochastic trajectory. Then, by
repeating it $\Sigma$ times, the stochastic average of any function can be
approximated by the arithmetic mean of the values of that function evaluated
at the different stochastic trajectories.


\begin{thebibliography}{99}
\bibitem {BTExperiments}We would like to note that experiments related to this
DOPO's configuration have been recently performed below threshold \cite{EXP}.
\bibitem {OPOnote}This result is linked to the noise reduction in the
signal-idler intensity difference of a non-degenerate optical parametric
oscillator \cite{OPO}. From a rigurous point of view, this is seen by writing
a representation of the signal field's OAM in terms of boson operators as
$\hat{L}=\int d^{2}\mathbf{r}\hat{A}_{\mathrm{s}}^{\dagger}\left(
\mathbf{r}\right)  \left(  -i\partial_{\phi}\right)  \hat{A}_{\mathrm{s}%
}\left(  \mathbf{r}\right)  =\hat{a}_{+1}^{\dagger}\hat{a}_{+1}-\hat{a}%
_{-1}^{\dagger}\hat{a}_{-1}$ -see Eq. (\ref{SVEs})-, showing that noise
reduction in the intensity difference between opposite OAM\ modes should
entail squeezing in the OAM of the generated pattern.
\bibitem {CompDifPRL}Note that the quantitative difference between this
difusion coeficient $D$ and that of \cite{PRL}, named $D_{\theta}$ in that
reference, comes from the fact that $D$ is $D_{\theta}$ normalized to
$\gamma_{\mathrm{s}}$.
\bibitem {DOPOspectrum}In the single-mode DOPO the best levels of squeezing
are found below threshold (within the linearized theory), where its noise spectrum can be written as
$V^{\mathrm{out}}\left(  \omega\right)  =1+S_{+}\left(  \omega\right)
\cos^{2}\varphi-S_{-}\left(  \omega\right)  \sin^{2}\varphi$, with $S_{\pm
}\left(  \omega\right)  =4\sigma/\left[  \left(  \sigma\pm1\right)
^{2}+\omega^{2}\right]  $.
\bibitem {Biorthonormal}This means that there exist a set of eigenvectors
$\mathbf{v}_{m}$ satisfying $\mathcal{L}\mathbf{v}_{m}=\lambda_{m}%
\mathbf{v}_{m}$, and another set $\mathbf{w}_{m}$ satisfying $\mathcal{L}%
^{\dagger}\mathbf{w}_{m}=\lambda_{m}^{\ast}\mathbf{w}_{m}$, so that
$\mathbf{w}_{m}^{\ast}\cdot\mathbf{v}_{m}=\delta_{mn}$.
\end{thebibliography}
\end{document}